\documentclass{aa}
\newcommand{\Bildbreite}[1]{#1}

\usepackage{color,xcolor}

\usepackage{hyperref}
\usepackage{txfonts}
\usepackage{grffile}

\newcommand {\Msun}{M_\odot} %
\def\Msol{M_\odot} %
\newcommand{\Rsun}{R_\odot}

\newcommand {\Lsun}{L_\odot}

\newcommand{\hd}{HIP\,65426}
\newcommand{\hdb}{HIP\,65426\,b}
\newcommand{\hip}{HIP\,65426}
\newcommand{\hipb}{HIP\,65426\,b}
\newcommand{\MJ}{{M_{\textnormal{J}}}}
\newcommand{\RJ}{{R_{\textnormal{J}}}}

\newcommand{\eqsep}{\;\;\;\;}
\newcommand{\me}{\, {\rm M}_{\oplus}}

\def\rp{{a_{\rm p}}}
\def\mp{{m_{\rm p}}}
\def\ep{{e_{\rm p}}}

\newcommand{\tauR}{{\tau_{\mathrm{R}}}}

\def\ej{_{\rm ej}}
\def\tide{_{\rm tide}}

\def\d{{\rm d}}

\def\tKH{t_{\textrm{KH}}}
\def\tKuehl{t_{\rm cool}}

\def\DtForm{{\Delta t_{\rm form}}}

\def\rhogas{\rho_{\mathrm{gas}}}

\def\Mstar{m_\star}
\def\vstar{v_\star}

\newcommand{\kB}{k_{\textrm{B}}}
\newcommand{\spf}{s_{\textrm{pf}}}
\newcommand{\spfmin}{s_{\textrm{pf, min}}}
\newcommand{\spfmax}{s_{\textrm{pf, max}}}
\newcommand{\spfminhot}{\spfmin^{\textrm{hot}}}
\newcommand{\spfmaxhot}{\spfmax^{\textrm{hot}}}
\newcommand{\spfmincold}{\spfmin^{\textrm{cold}}}
\newcommand{\spfmaxcold}{\spfmax^{\textrm{cold}}}

\def\lpf{L_{\rm pf}}

\usepackage{etoolbox}

\makeatletter
\patchcmd{\NAT@citex}
  {\@citea\NAT@hyper@{%
     \NAT@nmfmt{\NAT@nm}%
     \hyper@natlinkbreak{\NAT@aysep\NAT@spacechar}{\@citeb\@extra@b@citeb}%
     \NAT@date}}
  {\@citea\NAT@nmfmt{\NAT@nm}%
   \NAT@aysep\NAT@spacechar\NAT@hyper@{\NAT@date}}{}{}

\patchcmd{\NAT@citex}
  {\@citea\NAT@hyper@{%
     \NAT@nmfmt{\NAT@nm}%
     \hyper@natlinkbreak{\NAT@spacechar\NAT@@open\if*#1*\else#1\NAT@spacechar\fi}%
       {\@citeb\@extra@b@citeb}%
     \NAT@date}}
  {\@citea\NAT@nmfmt{\NAT@nm}%
   \NAT@spacechar\NAT@@open\if*#1*\else#1\NAT@spacechar\fi\NAT@hyper@{\NAT@date}}
  {}{}

\makeatother

\begin{document}

\title{Exploring the formation by core accretion\\ and the luminosity evolution
  of directly imaged planets}
\subtitle{The case of HIP\,65426\,b}
\titlerunning{Core accretion for directly imaged planets: HIP\,65426\,b}

\author{
Gabriel-Dominique Marleau\thanks{E-mail address: \texttt{gabriel.marleau\{@space.unibe.ch,} \texttt{@uni-tuebingen.de\}}.
Current affiliation: Institut f\"ur Astronomie und Astrophysik, Eberhard Karls Universit\"at T\"ubingen, Auf der Morgenstelle 10, 72076 T\"ubingen, Germany.}
\and Gavin A.\ L.\ Coleman
\and Adrien Leleu\thanks{CHEOPS Fellow}
\and Christoph Mordasini%
}
\authorrunning{Marleau et al.}

\institute{
Physikalisches Institut, Universit\"at Bern, Gesellschaftsstr.\ 6, 3012 Bern, Switzerland
}

\abstract
    { A low-mass companion to the two-solar mass star \hip\ has recently been detected
      by SPHERE at around 100~au from its host.
      Explaining the presence of super-Jovian planets at large separations,
      as revealed by direct imaging, is currently an open question.
    }
    {
      We want to derive statistical constraints on the mass and initial entropy of \hipb\ and to explore possible formation pathways
      of directly imaged  objects within the core-accretion paradigm, focusing on \hipb.
    }
    {
      Constraints on the planet's mass and post-formation entropy
      are derived from its age and luminosity combined with cooling models.
      For the first time, the results of population synthesis are also used to inform
      the results.
      Then a formation model that includes $N$-body dynamics
      with several embryos per disc is used to study possible formation histories and the properties
      of possible additional companions.
      Finally, the outcomes of two- and three-planet scattering
      in the post-disc phase are analysed, taking tides into account for small-pericentre orbits.
    }
    {
      The mass of \hipb\ is found to be
      $\mp=9.9^{+1.1}_{-1.8}~\MJ$  %
      using the hot population and
      $\mp=10.9^{+ 1.4}_{- 2.0}~\MJ$ %
      with the cold-nominal population.
      We find that core formation at small separations from the star followed by outward scattering
      and runaway accretion at a few hundred  astronomical units
      succeeds in reproducing the mass and separation of \hipb.
      Alternatively, systems having two or more giant planets close enough to be on an unstable orbit
      at disc dispersal are likely to end up with one planet on a wide \hipb-like orbit
      with a relatively high eccentricity ($\gtrsim 0.5$).
    }
    {
      If this scattering scenario explains its formation,
      \hipb\ is predicted to have a high eccentricity
      and to be accompanied by one or several roughly Jovian-mass planets
      at smaller semi-major  axes, which also could have a high eccentricity.
      This could be tested by further direct-imaging as well as radial-velocity observations.
    }

\keywords{planets and satellites: formation -- planet-disk interactions -- planets and satellites: dynamical evolution and stability
-- planets and satellites: physical evolution -- planets and satellites: individual: HIP~65426~b}

\maketitle

\section{Introduction}

In the last decade, direct imaging efforts have revealed a population of
super-Jovian planets at large separations from their host stars.
It has been well established  that these planets
are rare;  only a small percentage of stars 
possess such a companion \citep{bowler16}. %
What is not yet clear  is whether the formation process is intrinsically
inefficient there and how important post-formation architectural changes
to the system (through migration or interactions between protoplanets) are.
The formation mechanism that  produces these planets
has %
not yet been convincingly identified.
The main contenders are the different flavours of core accretion (CA;
with planetesimals or pebbles building up the core)
and of gravitational instability (GI; with or without tidal stripping).
Therefore, given the current low numbers of detections, %
every new data point can represent an important new challenge
for planet formation.

The first discovery of the SPHERE instrument at the VLT \citep{beuzit08,beuzit19},
\object{\hipb}, is an $\mp=8$--12~$\MJ$ dusty L$6\pm1$ companion to the $\Mstar=1.96\pm0.04~\Msol$ fast rotator \hip,
which has an equatorial velocity $\vstar\sin i=299\pm9$~km\,s$^{-1}$.
Its projected separation is $92.0\pm0.2$~au, and  %
the star is seen close to pole-on \citep{chauvin17}. If the planet is not captured
and its orbital plane is the same as the midplane of the star, the projected separation
is very close to the true separation. %

In this paper, we set out to explore how core accretion could lead
to the objects observed in direct imaging.
We take a closer look at \hipb\ because it is of low mass
and is at a relatively large separation, while its host star is a fast rotator.
Essentially, we are following up on the comment in \citet{chauvin17} that the `planet
location would not favor a formation by core accretion unless
\hipb\ formed significantly closer to the star followed by
a planet--planet scattering event.'

This paper is structured as follows.
In Section~\ref{Theil:MSi} we use planet evolution models and work backwards from the observations to derive
joint constraints on the mass and initial (i.e.\ post-formation) entropy of \hipb,
where `initial' refers to the beginning of the cooling.
We then switch to a forward approach and study the possible formation of \hipb.
In Section~\ref{sec:formationscenario} we use detailed planet formation models
following the disc evolution and $N$-body interactions.
Then in Section~\ref{sec:Gps} we use $N$-body integrations to look in detail at interactions between several companions
once the disc has cleared.
Finally, in Section~\ref{Theil:Zus} we present our conclusions and a discussion.

\section{Constraints on the mass and post-formation entropy}
 \label{Theil:MSi}

In this section we use the luminosity to derive, with planet evolution models, constraints on the mass
and initial (post-formation) entropy of \hipb,
following the approach of \citet{mc14}, as also applied
to $\kappa$~And~b and $\beta$~Pic~b \citep{bonnefoy14kap,bonnefoy14betaPic}.
The idea is to explore the parameter space of mass and initial entropy
through the Markov chain Monte Carlo (MCMC) method.
We assume Gaussian error bars on the logarithm of the luminosity
and on the linear age.
Both flat and non-flat priors in linear mass and post-formation
entropy are considered, as detailed in Section~\ref{Theil:M,Si-prior}.

\subsection{Luminosity and age}
 \label{Theil:L-Kurven}
Firstly, we discuss the input quantities for the MCMC.
The adopted bolometric luminosity is $\log L/\Lsun =-4.06 \pm 0.10$ as derived by \citet{chauvin17}.
Contrary to estimates based on the photometry in individual bands,
this quantity should be robust as it is based on the comparison to young L5--L7 dwarfs
with a similar near-infrared spectrum.

The typical age of stars in the Lower Centaurus--Crux group around \hip\ is $14\pm2$~Myr, but
the placement of phase-space neighbours of \hip\ in a Hertzsprung--Russell diagram 
suggests an age of 9--10~Myr.
This lead \citet{chauvin17} to adopt an age of $14\pm4$~Myr.
We note that at 2~$\Msol$, \hip\ is predicted by stellar evolution models to have
a pre-main sequence lifetime of approximately 15~Myr (see the overview as a function
of stellar mass in fig.~1 of \citealp{dotter16}),
so that it is approaching the main sequence or has only recently joined  it.
A point to consider is that
if \hipb\ formed by core accretion (CA), its cooling age would be smaller
by a not entirely negligible formation delay $\DtForm$ \citep{fortney05,bonnefoy14betaPic},
which we now briefly discuss.

While the dependence of the formation time $\DtForm$ on stellar mass
has not yet been studied in detail,
it seems plausible that giants form more quickly around more massive stars.
Since runaway accretion proceeds very quickly by construction,
it is the oligarchic growth phase that dominates the total formation time.
For instance, \citet[][their eq.~(11)]{thommes03} found that in this regime,
the planet growth rate scales as $\dot{M} \propto \Mstar^{1/6} \Sigma_m\rhogas^{2/5}$,
where $\Sigma_m$ and $\rhogas$ are respectively the surface density of planetesimals
and the (midplane) gas density. This scaling reflects in part the fact that
the core accretion rate is proportional to the Keplerian frequency, which at fixed orbital distance increases with stellar mass.  %
Since both $\Sigma_m$ and $\rhogas$ are expected to increase with stellar mass,
the formation time should decrease with planet mass.
Also, observationally, the formation time $\DtForm$ is unlikely much longer than 3~Myr
since discs around more massive stars are shorter-lived \citep{kenken09,ribas15};
already at solar masses,
gas giants must form typically in at most $\DtForm\sim3$--5~Myr
given the lifetimes of protoplanetary discs \citep{haisch01}.
Finally,
population synthesis calculations for a 2~$\Msol$ central star (Mordasini et al., in prep.) %
indicate that most $\sim10~\MJ$ planets (approximately the mass of \ \hipb, as we show later)
have reached their final mass after roughly $\DtForm\sim 2$~Myr, and
the simulations presented in Section~\ref{sec:formationscenario} using the \citet{ColemanNelson16b} models
for a 2~$\Msol$ star yield $\DtForm\approx 2.5$--4~Myr.
Therefore, we adopt  $\DtForm=2$~Myr, and thus $\tKuehl = 12\pm4$~Myr as the fiducial age.
We note that this $\DtForm$ is of the order of or smaller than the one-sigma error bar  on the age.
However, to  address  formation by gravitational instability, where we expect the planet
to be approximately coeval with the star, we also study the case of $\tKuehl=14\pm4$~Myr.

\subsection{BEX cooling curves}
 \label{Theil:BEX-Modelle}

For the MCMC we use the Bern EXoplanet cooling curves (BEX)
with the AMES-COND atmospheres.
The BEX models use the Bern planet evolution (cooling) code
\texttt{completo~21}, which includes the cooling and contraction
of the core and envelope at constant mass
(see sects.~3.2 and~3.8.3 of \citealt{morda12_I},
sect.~2.3 of \citealt{morda12_II}, and sect.~2 of \citealt{linder19})
as well as deuterium burning \citep{moll12}.
The boundary conditions are provided by atmospheric models.
Previously,
only the simple Eddington model had been implemented,
but we can now use arbitrary atmospheric models,
following the coupling approach of \citet{chabrier97}.
This entails simply taking a pressure--temperature point in the adiabatic part
of the deep atmosphere as the starting point of the interior structure calculation.
Since the structure is adiabatic, the precise location (e.g.\ at a Rosseland optical
depth $\Delta\tauR=100$, at a pressure $P=50$~bar,
or at the top of deepest convection zone)
will not matter, and it is easy to verify that in any case the error
in the radius is at most of a few percentage points.
This coupling approach was applied recently to low-mass planets in \citet{linder19}.

\begin{figure}[tb]
\centering
\includegraphics[width=\Bildbreite{0.47}\textwidth]{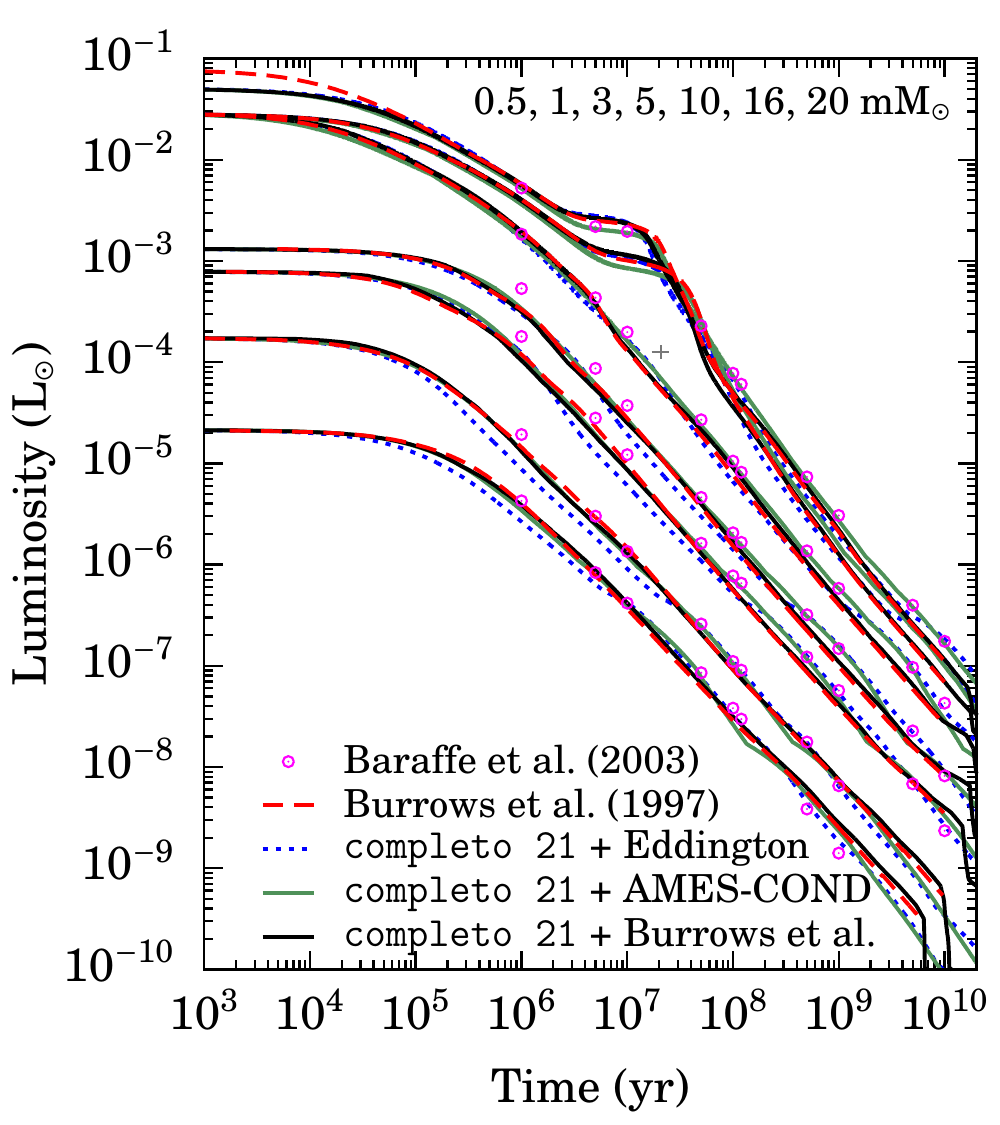}
\caption{
\label{Abb:c21-Kurven}
Bern EXoplanet cooling curves (BEX) for planet masses $\mp=0.5$--20~m$\Msol$ (bottom to top)
with different atmospheric boundary conditions (Eddington, \citealp{baraffe03},
\citealp{burr97}; see legend).
The Bern evolution (cooling) code \texttt{completo~21} is used
and compares very well to the original models.
Units of milli-solar masses (1~m$\Msol=1.05~\MJ$) are used to reproduce
as closely as possible the tracks of \citet{burr97} and \citet{baraffe03}.
The starting luminosities of the original AMES-COND \protect\citep{baraffe03} tracks
are apparently not quite the same as for \protect\citet{burr97}.
The faint grey cross shows $\beta$~Pic~b \citep{bonnefoy14betaPic}
as an example error bar.
}
\end{figure}

Currently, the BEX models are available with boundary conditions
provided by
\begin{enumerate}
\item[(i)]~the Eddington assumption;
\item[(ii)]~AMES-COND \citep{baraffe03};
\item[(iii)]~\citet{burr97};
\item[(iv)]~\texttt{petitCODE} \citep{moll15}; 
\item[(v)]~\texttt{HELIOS} (\citealp{malik17}, Malik et al., in review).
\end{enumerate}
For flavours~(ii) and~(iii) we extracted the relevant information
from the publicly available \citet{burr97} tracks %
and the \citet{baraffe03} grids\footnote{See \url{https://www.astro.princeton.edu/~burrows/dat-html/data/} and \url{https://phoenix.ens-lyon.fr/Grids/AMES-Cond/STRUCTURES/}, respectively.}.
The details will be described in a dedicated publication,
but we note already that
we can reproduce very well  the \citet{burr97}
and the AMES-COND tracks (see Fig.~\ref{Abb:c21-Kurven}).

By default, the BEX curves
assume full ISM deuterium abundance at the beginning of cooling,
while in fact in core accretion a mass-dependent fraction will be burnt during formation \citep{moll12,3M}.
However, in both cold 1~$\Msol$ and 2~$\Msol$ population syntheses, 
objects need a mass of 16~$\MJ$ (20~$\MJ$) to have consumed even only $\approx30$\,\%\ ($\approx70$\,\%)
of their initial D abundance by the end of formation.
Given the masses we  find later for \hipb,
and since in GI it is likely that
no deuterium is destroyed during formation,
the use of full deuterium abundance at the beginning of cooling is inconsequential.

We display in Fig.~\ref{Abb:c21-Kurven} the different flavours of the
BEX cooling curves compared to the classical models of \citet{burr97} and \citet{baraffe03}.
The initial luminosities are set to the same values as in \citet{burr97},
except for the 20~m$\Msol$ case for which we took a slightly lower initial luminosity
to avoid non-monotonicities in the re-interpolation of the original \citet{burr97} data.
Otherwise, the BEX curves clearly follow the \citet{burr97} models, including the `shoulder'
that occurs during deuterium burning.
At very old ages (20~Gyr) the black lines diverge because the models are beyond the tabulated range of input atmospheric structures.

We also see that the choice of either of the three classic atmospheric models (Eddington, AMES-Cond, Burrows)
as outer boundary conditions only has a small effect on the cooling,
as expected \citep{baraffe03,chabrier00evol}.
Furthermore, it should be noted that the starting luminosities of the original
AMES-COND \citep{baraffe03} tracks
are apparently not quite the same as for \citet{burr97}.

\subsection{First analysis of the mass}

\begin{figure*}[tb]
  \centering
\includegraphics[width=0.49\textwidth]{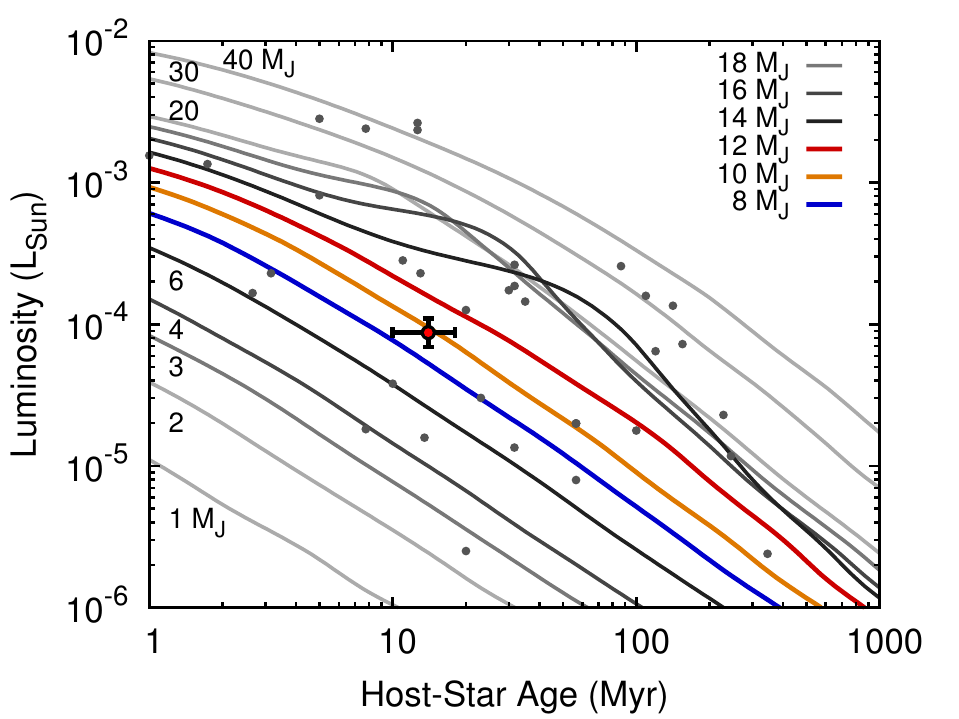}~\includegraphics[width=0.49\textwidth]{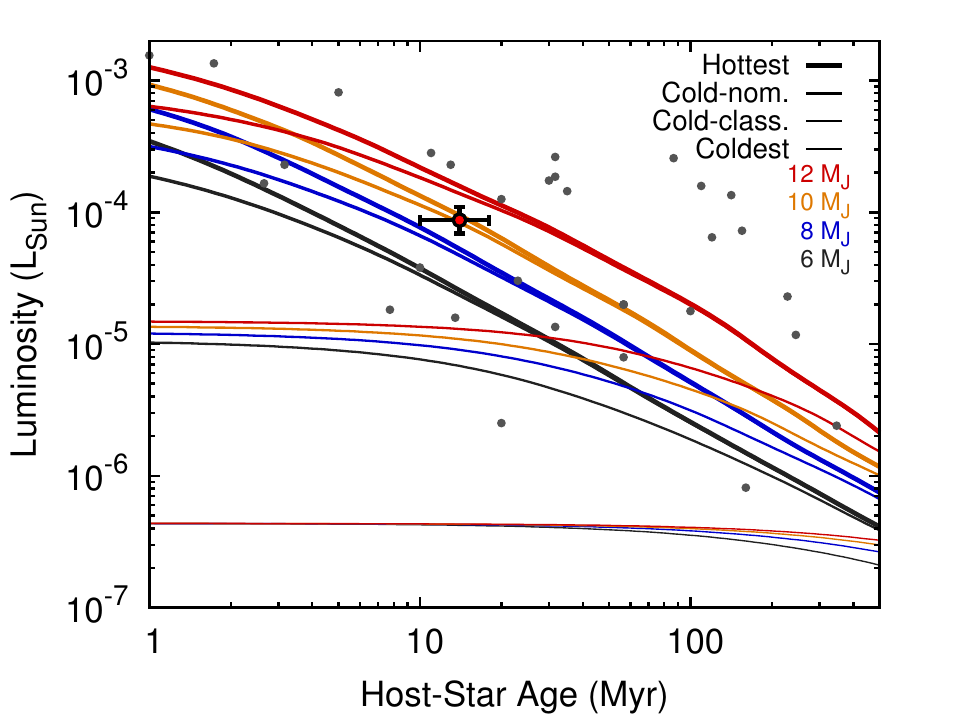}
\caption{
\label{Abb:Lt}
\textit{Left panel:}
Placement of \hipb\ (point with error bars) in the age--luminosity diagram.
The dots show other direct detections from the literature; the  error bars are omitted for clarity.
No formation delays $\DtForm$ are subtracted.
The cooling curves are the BEX hottest starts (Eq.~\ref{Gl:lpf hottest})
with the AMES-COND \citep{baraffe03} atmospheres
for masses of $\mp=1$--40 $\MJ$ (bottom to top; see labels and legend).
\textit{Right panel:}
Effect of
different post-formation luminosities, as given by the populations
of \citet{3M}:
hottest starts (as in the left panel),  %
cold-nominal population, cold-classical population,
and coldest starts (thick to thin lines; see Eq.~(\ref{Gl:lpf})).
Only masses of $\mp=6$~(black), 8~(blue), 10~(orange), and $12~\MJ$ (red)
are shown (bottom to top).
The  axis ranges relative to the left panel are different.
}
\end{figure*}

Figure~\ref{Abb:Lt}a compares \hipb\ to direct detections
and the `hottest start' cooling tracks of \citet{3M},
which use the simple Eddington outer boundary condition.
A direct comparison with these cooling curves suggests a mass $\mp\approx8$--11~$\MJ$,
which is not rare for direct detections of young isolated brown dwarfs (in the sense of substellar-mass objects;
see the  mass histogram in fig.~18 of \citealt{gagn15}).
As shown in Fig.~\ref{Abb:c21-Kurven}, these simpler models
quite closely match  cooling tracks based on detailed atmospheric
models such as \citet{burr97} or \citet{baraffe03}.
However, the luminosity error bar (0.1~dex, also a typical size; see  e.g.\ \citealp{bowler16})
is small enough for the derived mass to depend slightly on the choice of the cooling curves.

After this first estimate of the mass based on models
with an arbitrarily high post-formation luminosity,
we look at cooling curves whose post-formation (also termed initial) luminosity $\lpf$
follows the four relations seen in the population syntheses of \citet[their sect.~5.2.2 and their fig.~13]{3M}.
For $\mp\approx0.3$ to $\approx12$~$\MJ$ (i.e.\ for planets
that are massive enough to undergo the detached phase during the presence of the nebula,
but not massive enough for deuterium burning to occur), these relations are given by
\begin{subequations}
\label{Gl:lpf}
\begin{align}
\lpf^{\textrm{hottest}} &= 7.3\times10^{-5}\Lsun\, (\mp/\MJ)^{1.4} \label{Gl:lpf hottest},\\
\lpf^{\textrm{cold-nom.}} &= 2.6\times10^{-5}\Lsun\, (\mp/\MJ)^{1.3}\label{Gl:lpf cold-nom.},\\
\lpf^{\textrm{cold-class.}} &= 4.3\times10^{-6}\Lsun\, (\mp/\MJ)^{0.5},\\
\lpf^{\textrm{coldest}} &= 4.3\times10^{-7}\Lsun,
\end{align}
\end{subequations}
respectively, for the hottest, cold-nominal, cold-classical, and coldest planets.
Briefly, $\lpf^{\textrm{hottest}}$ traces the brightest planet at every mass;
$\lpf^{\textrm{cold-nom.}}$ 
corresponds
to the cold-nominal population,
in which gas is assumed to accrete cold;
$\lpf^{\textrm{cold-class.}}$ is the best fit to the cold-classical population
(which however shows an appreciable spread in luminosity at a given mass),
in which the core artificially stops growing in the runaway phase \`a la \citet{marley07}; and finally,
$\lpf^{\textrm{coldest}}$ traces the coldest planets at a given mass, which come from the small-core (coldest-start) population.
It should be noted that we defined here the cold-nominal relation (Eq.~(\ref{Gl:lpf cold-nom.})) not as the mean of the cold-nominal population (as in \citealt{3M}, with $\lpf=1.2\times10^{-5}\Lsun\, (\mp/\MJ)^{1.3}$), but as the approximate upper envelope of points of that population.

Cooling tracks from all four relations are shown in Fig.~\ref{Abb:Lt}b.
At this age and for this mass there is barely any difference in the cooling curves
of the hottest and the cold-nominal starts.
However, the luminosities in the cold-classical population are one order of magnitude lower, the initial cooling (Kelvin--Helmholtz) timescale
being $\tKH\sim100~\textrm{Myr}$, which is roughly ten times longer
than the age of \hipb.
The coldest starts, finally, are even several orders of magnitude fainter than the others,
with at the lower masses an initial $\tKH\sim500$~Myr.
Since the initial luminosities are well below the observed luminosity,
no coldest start can match the observed luminosity of \hipb.
Deuterium burning in more massive objects would be required here to reproduce the observed luminosity.
In any case, as argued from different points of view by \citet{3M} and \citet{berardo17},
the coldest starts are not expected to be realistic.

We conclude that in this simple analysis, only the hottest and cold-nominal populations can reproduce \hipb. In the next section, we revisit this analysis in a more systematic fashion and take the error bars on age and luminosity into account.

\subsection{Inputs: priors on mass and luminosity}
 \label{Theil:M,Si-prior}

Recently, using the tool of population synthesis, \citet{3M} presented the first discussion
of the statistics of planetary luminosities
as predicted by a planet formation model.
They looked in particular at the core accretion paradigm \citep{pollack96,morda12_I} and considered three populations,
differing in the assumed efficiencies of the accretional heating of gas and
planetesimals during formation:
\begin{enumerate}
 \item[(i)] a cold-nominal population, in which the entire gas accretion luminosity is radiated away at the shock,
            as in \citet{morda12_II};
 \item[(ii)] a hot population, which differs from the first only by the assumption that
             the entire accretion luminosity is brought into the planet; 
 \item[(iii)] a cold-classical population, which assumes, as in the classical work by \citet{marley07}, that
planetesimal accretion stops
artificially once a giant planet enters the disc-limited gas
accretion (detached) phase,
and also does not include planetary migration.
\end{enumerate}
Since the cold-classical population serves rather for model comparisons,
and given that first dedicated and systematic simulations of the accretion shock have been recently
performed (\citealp{mkkm17}; Marleau et al., in prep.) but not yet used to produce cooling curves,
we  consider in this work the cold-nominal and hot populations as
more realistic extreme scenarios.

We now turn to the total distribution function, %
which we write as
\begin{equation}
 \label{Gl:Prior beide}
 \frac{\d^2 N}{\d \mp\,\d\spf} = p(\mp,\spf) = p_{\spf}(\mp,\spf)\times p_\mp(\mp).
\end{equation}
\citet{3M} showed that there is spread of post-formation entropies
of approximately $\Delta\spf\approx1~\kB\,\textrm{baryon}^{-1}$ at a given mass
(see their fig.~12), coming mostly from the core-mass effect \citep{morda13,bodenheimer13}.
Given that the distribution of entropies is rather uniform for a given mass,
we fit simple mass-dependent top-hat functions
to the probability distributions of $\spf$:
\begin{align}
 \label{Gl:SiPrior}
 p_{\spf}(\mp,\spf) = \left\{
       \begin{array}{l} 
         1  \eqsep\mbox{if $\spfmin(\mp) < \spf < \spfmax(\mp)$}\\  %
         0  \eqsep\mbox{otherwise}.
       \end{array}
       \right.
\end{align}
The following functions, dropping the usual entropy units $\kB\,\textrm{baryon}^{-1}$,
closely fit  the envelope of points $\spf(\mp)$ in \citet{3M}.
For the cold-nominal population,
the lower and upper edges are given respectively by
\begin{subequations}
 \label{Gl:SiPrior kalt}
\begin{align}
 \spfmincold  &= \left\{
       \begin{array}{l}
         9.40 + 0.07 \,(\mp-13.6)     \eqsep\eqsep\mbox{if $2<\mp<13$}\\  %
        11.200 - 0.033 \,(\mp-20)^2   \eqsep\mbox{otherwise}
       \end{array}
       \right.\\
 \spfmaxcold &= 10.700 + 0.116 \, (\mp-10),
\end{align}
\end{subequations}
where masses $\mp$ are implicitly in Jupiter masses in these equations, while for the hot population,
the bounds are
\begin{subequations}
 \label{Gl:SiPrior warm}
\begin{align}
 \spfminhot &= 10.00 + 0.12 \,(\mp-10)\\  %
 \spfmaxhot &= 11.300 + 0.116 \,(\mp-10).
\end{align}
\end{subequations}
This holds down to 2~$\MJ$.
We point out that in the Bern planet formation code, as in most codes using the \citet{scvh} equation of state,
there is an entropy offset relative to the published \citet{scvh} tables (see Appendix~B in \citealp{mc14}).
This difference has no physical meaning, but care must be taken when comparing to work using
codes with other entropy reference points such as \texttt{MESA} \citep{paxton11,paxton13,paxton15} as used by \citet{berardo17}.

Marginalising over entropy,
the mass function in both the cold- and hot-start populations
is, for about 1 to 10~$\MJ$, approximately given by
\begin{equation}
 \label{Gl:MPrior}
  p_\mp(\mp) = \frac{\d N}{\d \mp}\propto \mp^{-1},
\end{equation}
i.e.\ the distribution is nearly flat in $\log \mp$.
As mentioned by \citet{3M}, %
 this is similar to the distribution found by \citet{morda09b} for population synthesis planets
detectable by radial velocity,
which in turn agreed with the $\d N/\d \mp\propto \mp^{-1.05}$ fit of \citet{marcy05}.
We note,  however, that \citet{cumming08} found $\d N/\d \mp\propto \mp^{-1.3\pm0.2}$
but for periods $<2000$~days,
while \citet{brandt14} obtained from direct imaging $\d N/\d \mp\propto \mp^{-0.7\pm0.6}$ at distances $\gtrsim10$~au.
Larger numbers of log-period radial velocity and direct-imaging detections will be necessary to reduce
the error bars on these exponents.

\subsection{Results: Mass--entropy constraints}

\begin{figure*}[tb]
  \centering
\includegraphics[width=\Bildbreite{0.8}\textwidth]{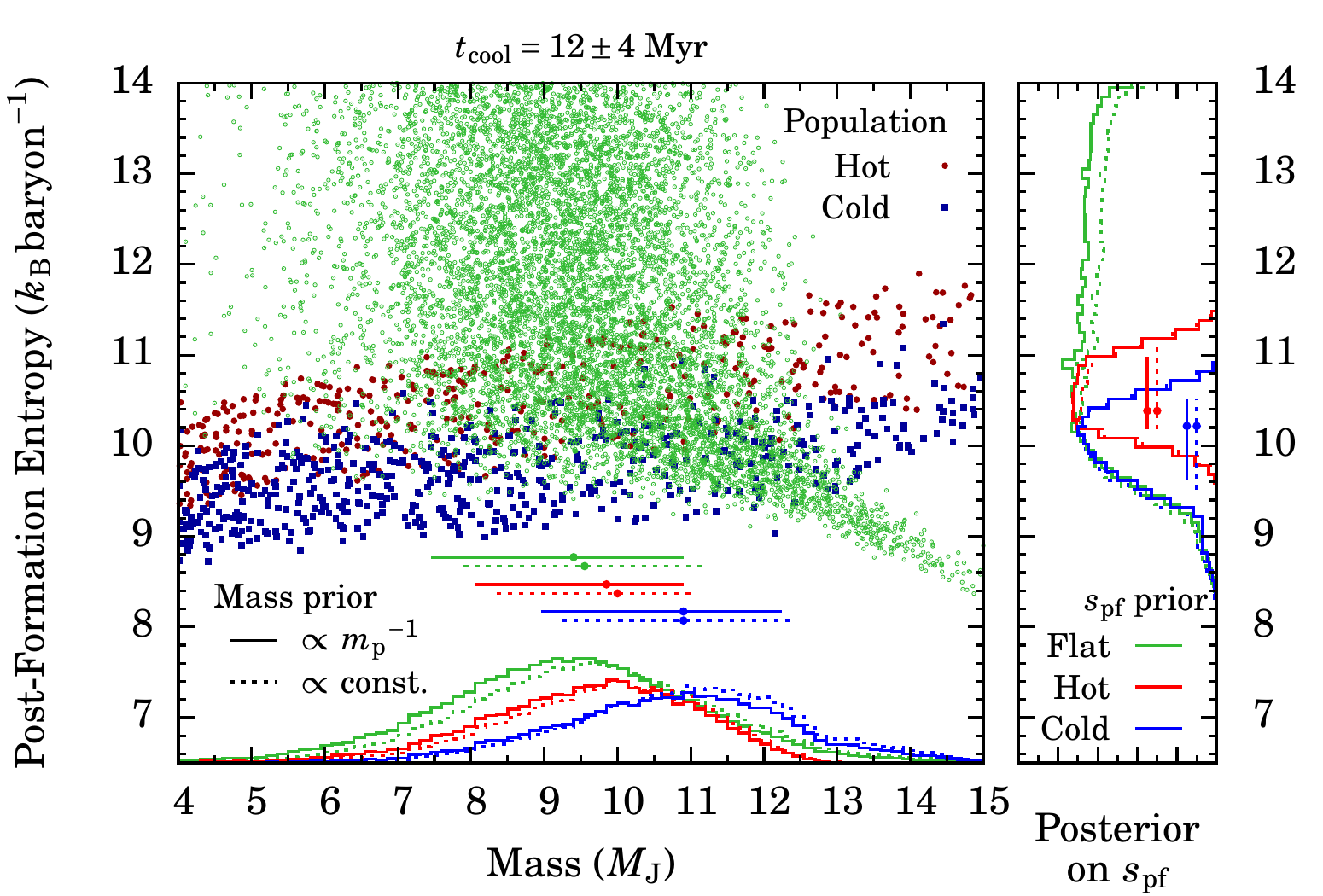}
\caption{
\label{Abb:MSi}
Statistical constraints on the mass and post-formation entropy of \hipb\ from its age and luminosity.
Green dots show the outcome of the MCMC using the
BEX models with the AMES-COND atmospheres (\textsection~\ref{Theil:BEX-Modelle}).
The cooling age is $\tKuehl=12\pm4$~Myr,  %
which is $\Delta t_{\textrm{form}}=2$~Myr less than the star's age,
and the luminosity is $\log L/\Lsun = -4.06\pm0.10$.
Results from the hot (cold) population syntheses of \citet{3M} are shown in dark red (dark blue).
Marginalised posteriors are displayed at the bottom and in the side panel:
with a flat prior, with the prior from the hot population (Eq.~(\ref{Gl:SiPrior warm})),
and with the prior from the cold population (Eq.~(\ref{Gl:SiPrior kalt}))
(green, red, and blue lines, respectively,  from top to bottom).
The full lines also use  the mass prior $\d N/\d \mp\propto \mp^{-1}$ (Eq.~(\ref{Gl:MPrior})),
whereas the dotted lines use a flat prior in mass.
The points with error bars show the corresponding peaks of the posteriors and the 68.3\,\%\ confidence intervals.
}
\end{figure*}

Figure~\ref{Abb:MSi} shows the joint constraints on the mass and post-formation (or initial) entropy %
using the different priors discussed above.
Considering first the case of uniform priors (i.e.\ not using information from formation scenarios),
we find that
the post-formation entropy $\spf\geqslant~9.2~\kB\,\textrm{baryon}^{-1}$
but it is not otherwise constrained.
This lower limit holds independently of the formation pathway
and for masses up to $\mp\approx15~\MJ$ (a conservative assumption).
Marginalising instead over entropy, the 68.3\%\ confidence interval (which is  used throughout this section despite the non-Gaussianity of the posteriors)
on the mass is
$\mp= 9.6 \pm1.7~\MJ$.  %
For high values of $\spf$,
the BEX models using the AMES-COND boundary conditions
closely match  the \citet{baraffe03} hot-start  cooling tracks for these masses.
If we consider somewhat arbitrarily $\spf\gtrsim14$ to approximate what  %
is usually thought of as hot starts, we find
$\mp= 9.0 ^{+ 1.3 }_{-1.5  }~\MJ$ for a cooling age $\tKuehl 12\pm4$~Myr. 
This agrees well with the $\mp=10\pm2~\MJ$ reported by \citet{chauvin17} for the DUSTY models\footnote{
  The value $\mp=7_{-1}^{+2}~\MJ$ quoted by \citet{chauvin17} for COND03 \citep{baraffe03}
  does not come from a luminosity comparison and is therefore less robust.
  However, COND03 and DUSTY use by construction the same luminosity tracks \citep{baraffe03}.}.
As expected from \citet{mc14}, the relative uncertainty on the hot-start mass
$\sigma_\mp/\mp\approx0.2$ is $\approx\frac{1}{2}\sigma_{\tKuehl}/\tKuehl\approx 0.3$.

Next we fold in the outcome of the population syntheses into the analysis.
If we take only the mass prior (Eq.~(\ref{Gl:MPrior})) into account,
we obtain
$\mp=9.4^{+ 1.5 }_{- 2.0 }~\MJ$,
which is lower by $\Delta\mp\approx0.2~\MJ$ than the case without priors.  %
Applying  the $\spf$ priors (Eq.~(\ref{Gl:SiPrior})) as well, %
we obtain
$\mp=10.9^{+ 1.4}_{- 2.0}~\MJ$  %
for the cold population (Eq.~(\ref{Gl:SiPrior kalt}))
and
$\mp=9.9^{+1.1}_{-1.8}~\MJ$   %
for the hot population (Eq.~(\ref{Gl:SiPrior warm})).

These masses are shown as points with horizontal bars in the main panel of Fig.~\ref{Abb:MSi}.
The difference between the mass inferred with and without the mass priors is small,
with $\Delta\mp\lesssim0.2~\MJ$.  %
These differences represent only a modest fraction of the error bars.
However, the $\spf$ priors are mildly important, leading to a difference
$\Delta \mp\approx1~\MJ$  %
between the hot and the cold populations and even $\Delta \mp=1.5~\MJ$ between the flat-prior
and (with Eq.~(\ref{Gl:Prior beide})) the cold-population cases.  %
Finally, we note the distinctly asymmetrical shape of the confidence intervals
when using the priors.
This asymmetry comes mostly from the $\spf$ prior despite the $p_\mp \propto \mp^{-1}$ scaling
since the mass interval is small.

The posterior on the post-formation entropy changes dramatically
when taking the population-synthesis priors into account,
as visible in the right panel of Fig.~\ref{Abb:MSi}.
The lower bound $\spf\gtrsim9.2$ obtained with the uniform prior
does not change, but the population-synthesis priors lead
to the determination of an upper bound, yielding
$\spf=10.4  ^{+ 0.7 }_{- 0.2 }$ %
in the case of the  hot population  and
$\spf=10.2  ^{+ 0.3 }_{- 0.7 }$  %
for the cold.
It should be noted that the probability maxima are rather flat.
These values differ only marginally from each other,
reflecting the large overlap between the post-formation entropies or luminosities
of the cold- and hot-start populations, which is ultimately a consequence
of the core-mass effect (CME) as discussed by \citet[][sect.~5.2.1]{3M}.
These values $\spf\approx10.3$ are clearly lower than classical (arbitrarily) hot starts ($\spf\approx13$),
with an initial Kelvin--Helmholtz time $\tKH\sim10$~Myr as opposed to $\tKH\lesssim1$~Myr
for classical hot starts. Thus, \hipb\ would have just begun joining the hot-start cooling track
(see \citealp{mc14} for a general discussion of the shape of cooling tracks).
We finally note that the mass prior barely changes the $\spf$ posteriors.

\subsection{Discussion}

For comparison, with a shorter cooling age $\tKuehl=10\pm4$~Myr
(i.e.\ coming from a longer formation period),
we obtain with uniform priors
$\mp=9.0^{+1.9}_{-1.7}~\MJ$  %
and with only the mass prior
$\mp=8.7  ^{+ 1.9 }_{- 2.0 }~\MJ$,  %
whereas using the mass and $\spf$ priors from hot-start (cold-start) populations yields
$\mp=9.3  ^{+ 1.3 }_{- 1.8 }~\MJ$ %
($\mp = 10.3  ^{+ 1.6 }_{- 1.8 }~\MJ$). %
Instead, taking a cooling age  $\tKuehl=14\pm4$~Myr, i.e. the age of \hip\ as might
correspond to formation by gravitational instability,
with
only mass priors we obtain
$\mp=  9.9 ^{+ 1.4 }_{- 1.7 }~\MJ$;
instead, using the $\spf$ and mass priors from the hot-start (cold-start)  populations yields
$\mp= 10.4 ^{+ 1.0 }_{- 1.6 }~\MJ$  %
($\mp=11.3 ^{+1.0 }_{-1.7 }~\MJ$).
This is somewhat higher than, but still consistent with, the mass
derived by \citet{cheetham19}. Using an age of $14\pm4$~Myr and the AMES-Cond models,
they found $\mp=7.5\pm0.9~\MJ$ based on magnitudes in individual bands
and $\mp=8.3\pm0.9~\MJ$ based on their bolometric luminosity,
which had an uncertainty $\sigma_{\log L}=0.03$~dex half as large
as the value used here. That the mass found by \citet{cheetham19} is lower than
that derived here is not surprising
since the AMES-Cond models they used correspond only to hotter starts,
whereas here a range of $\spf$ was considered.

In general, one could  expect somewhat different results
if using the logarithm of the post-formation luminosity
instead of the post-formation entropy as an independent variable
(along with the mass).
Indeed, the luminosity $L$ and entropy $s$ are monotonic functions
of each other at a given mass,
but the slope $\d\log L/\d s$  depends on both mass and entropy
\citep{mc14}.
This means that a prior which is uniform in $s$ for all masses
is not uniform in $\log L$ for all masses, and vice versa.

However, one can argue that this should be of negligible concern.
In the case of a flat prior in $\spf$,
the posterior  was also relatively flat, and a small distortion
will not change the nature of the weak constraints on $\spf$.
The distortion should be small judging by the precise scalings
identified in eq.~(9) of \citet{mc14},
and while these hold specifically for their Eddington atmospheric models,
the $L(s)$ relation will not be entirely different for AMES-COND. %
In the case of the hot or cold priors, the posteriors are non-zero
over  a relatively small region, so that in this case too there
should not be any significant skew.
  
We finally note that, as mentioned in Section~\ref{Theil:L-Kurven},
Fig.~\ref{Abb:MSi} shows that \hipb\ is unlikely to have a mass
for which a meaningful fraction of deuterium could be burnt (cf.\ \citealp{spiegel11,moll12}).
This justifies a posteriori the use of cooling curves
that assume full deuterium abundance at the start.
In the case of other detections close to the deuterium-burning limit
($\mp\approx13\pm2~\MJ$), however, this assumption would
need to be revisited if they formed over a longer timescale than
expected from gravitational instability.

To summarise, we find that the mass of \hipb\ is
$\mp=10.9^{+ 1.4}_{- 2.0}~\MJ$  %
with priors from the cold population and
$\mp=9.9^{+1.1}_{-1.8}~\MJ$
using the hot population.

\section{Forming HIP 65426 b in core-accretion models}
\label{sec:formationscenario}

We now switch from the study of \hipb's post-formation thermodynamical evolution to numerical experiments concerning its formation.
Core accretion models typically involve forming cores of giant planets and having them undergo runaway gas accretion
at small orbital radii ($\rp\lesssim20$~au) before they migrate in towards the central star, inconsistent with the location of \hdb.
At larger orbital radii, the time taken to form a core through planetesimal accretion is longer than typical protoplanetary disc lifetimes,
though forming a core through pebble accretion could be significantly faster (\citealp{Lambrechts14}; but see also \citealp{rosenthal18}).
Even if a single planetary core is able to form at large orbital radii, either through pebble or planetesimal accretion,
interactions with the local protoplanetary disc will force the planet to migrate through type I migration
to small orbital separations on timescales shorter than that required for the core to accrete
a significant gaseous envelope and undergo runaway gas accretion \citep{ColemanNelson16}.
This fast migration poses the main problem for forming a planet that has properties consistent with that found for \hdb.

To overcome these problems for the core accretion model in forming planets such as \hdb, we ran numerous $N$-body simulations
in which we placed a number of giant planet cores in a protoplanetary disc and allowed them to mutually interact,
migrate throughout the disc, and accrete gaseous disc material.
The idea is that as one giant planet core undergoes runaway gas accretion and rapidly increases its mass,
the system of planets becomes dynamically unstable, leading to the scattering of one of the less massive cores.
This core, once scattered out into the outer disc will then circularise its orbit and begin to migrate back in towards the central star.
However the core will continue accreting gas from the surrounding disc and could then undergo runaway gas accretion in the outer disc,
becoming a gas giant and transitioning to the slower type~II migration regime.
If the planet is scattered out far enough and has insufficient time to migrate back in towards the inner disc,
its final mass and semi-major axis could be similar to those of directly imaged planets and \hdb.
This process has been observed in population synthesis models when, in massive discs,
multiple gas giants form as a first generation and subsequently destabilise the orbits of surrounding embryos,
scattering them to larger orbits where they then grow into gas giants \citep{Ida2013}.

\subsection{Simulation set-up}
In order to run these simulations, we adapted the $N$-body and disc model of \citet{ColemanNelson16}
to be appropriate for a protoplanetary disc surrounding an A-type star such as \hd.
This model couples a 1D thermally evolving viscous disc model \citep{Shak} to the Mercury-6
symplectic integrator \citep{Chambers} and includes prescriptions for photoevaporation \citep{Dullemond,Alexander09},
both type I and II planet migration \citep{pdk11,LinPapaloizou86}, and gas accretion from the surrounding disc \citep{CPN17}.
Table~\ref{tab:nbodydisc} shows the disc parameters used for the simulations.
We chose the values for the viscosity parameter $\alpha$ and the photoevaporation factor $\Phi_{41}$
to give the disc an appropriate lifetime. Using the values presented in Table~\ref{tab:nbodydisc},
the initial disc had a total mass equivalent to $\sim8\%$
of the mass of \hd\ (i.e.~around $150~\MJ$) and a lifetime of 3.5~Myr.
The lifetime of the disc in the simulations is always shorter
since a significant fraction of the total gas mass is accreted onto the planets.

\begin{table}
\caption{Stellar and disc parameters used for the $N$-body simulations.
}
\label{tab:nbodydisc}
\centering
\begin{tabular}{ll}
\hline
\hline
Parameter & Value\\
\hline
Stellar mass & 2~$\Msun$ \\
Stellar radius & 2~$\Rsun$ \\
Stellar temperature & 10,000~K \\
Disc inner boundary & 0.1~au\\
Disc outer boundary & 200~au \\
Initial surface density exponent & $-1.5$\\
Initial surface density $\Sigma_{0}=\Sigma(1~\mathrm{au})$ & 8655~g\,cm$^{-2}$\\
Disc metallicity & 1 $\times$ solar \\
Photoevaporation factor $\Phi_{41}$ & 1000 \\
Background viscous $\alpha$ & $5\times 10^{-3}$ \\
\hline
\end{tabular}
\tablefoot{For the meaning of $\Phi_{41}$ see \citet{Dullemond}.
}
\end{table}

Since type I migration timescales for giant planet cores are shorter than the timescales
for the cores to reach runaway gas accretion \citep{ColemanNelson16}, we require a mechanism
to stall type I migration (see also \citealp{pudritz18}).
To stall type I migration and counter the short timescales experienced by giant planet cores,
we placed a radial structure in the disc that mimics the effects of a zonal flow.
Zonal flows have been observed in both local \citep{Johansen2009}
and global \citep{SteinackerPapaloizou2002,PapaloizouNelson2003,FromangNelson2006} numerical simulations
of magnetised discs, including those incorporating non-ideal MHD effects \citep{Bai2014,BethuneLesur2016,ZhuStoneBai2014}.
Radial structures which could also be reminiscent of zonal flows have also been seen
in numerous observations of protoplanetary discs \citep{HLTAU,Andrews16,VanBoekel17}.
The effect of zonal flows on a protoplanetary disc is to create a radial pressure bump
in the disc, which results in a positive surface density gradient.
This positive surface density gradient increases the strength of the vortensity component
of a planet's corotation torque, allowing it to balance the planet's Lindblad torque,
thus creating a planet trap that stalls type I migration \citep{Masset2006,Hasegawa11,ColemanNelson16b}.
To account for a radial structure in the disc that mimics the effects of a zonal flow,
we included a single radial structure in each simulation, following the approach used
in \citet[][see their section 2.3.3]{ColemanNelson16b}.
This radial structure increases the local $\alpha$ parameter when calculating the viscosity,
which results in a reduction in the local surface density, creating a positive surface density gradient
that acts as a planet trap as described above.
We assume that this structure remains at the same location in the disc, placed arbitrarily
at either 15 or 20~au in our simulations, and has a lifetime equivalent to the disc lifetime.
The lifetimes of zonal flows in MHD simulations are still unexplored
due to long simulation run times, but since these structures are seen in both young
and old protoplanetary disc observations \citep{HLTAU,Andrews16},
it seems reasonable to assume that the flows are long lived.

To account for planet migration we use the torque formulae of \citet{pdk10,pdk11} whilst the planet is embedded in the disc, to simulate type I migration due to Lindblad and corotation torques.
Our model accounts for the possible saturation of the corotation torque \citep{pdk11}, and also the influences of eccentricity and inclination on the disc forces \citep{cressnels,Fendyke}.
Once the planet has become massive enough to open a gap in the disc we use the impulse approximation to calculate the torques acting on the planet from the surrounding disc as it undergoes type II migration \citep{LinPapaloizou86}.
To calculate gas accretion on to the planet, we use the accretion routine presented in \citet{CPN17}.
In this model, whilst the planet is embedded in the disc we construct a 1D envelope structure model that self-consistently calculates the gas accretion rate taking into account local disc conditions.
After the planet has opened a gap in the disc, we assume that the gas accretion is equal to the viscous supply rate.
All gas that is accreted on to the planet is removed from the surrounding disc.

\begin{figure*}[th!]
\centering
\includegraphics[scale=0.6]{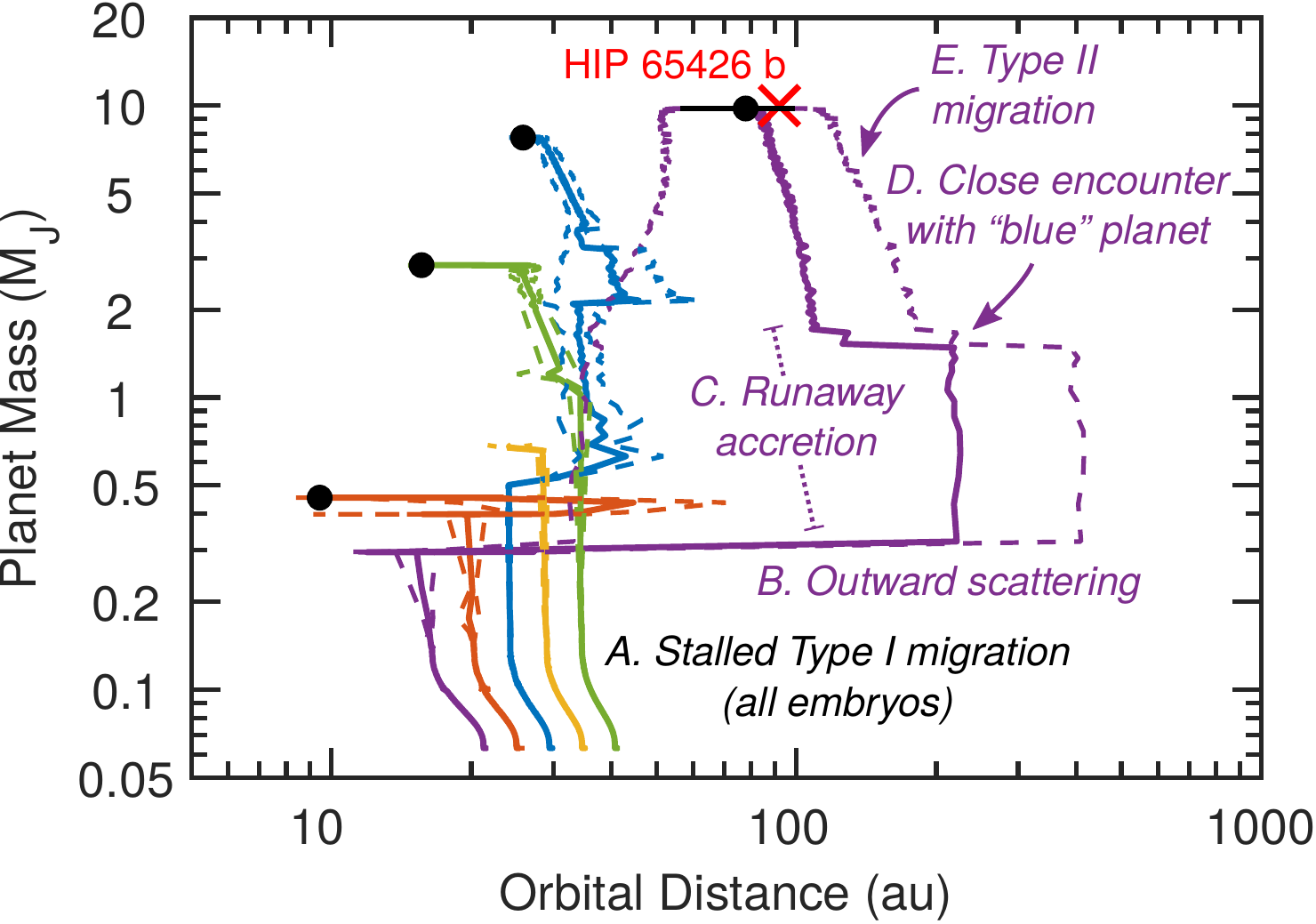}
\includegraphics[scale=0.6]{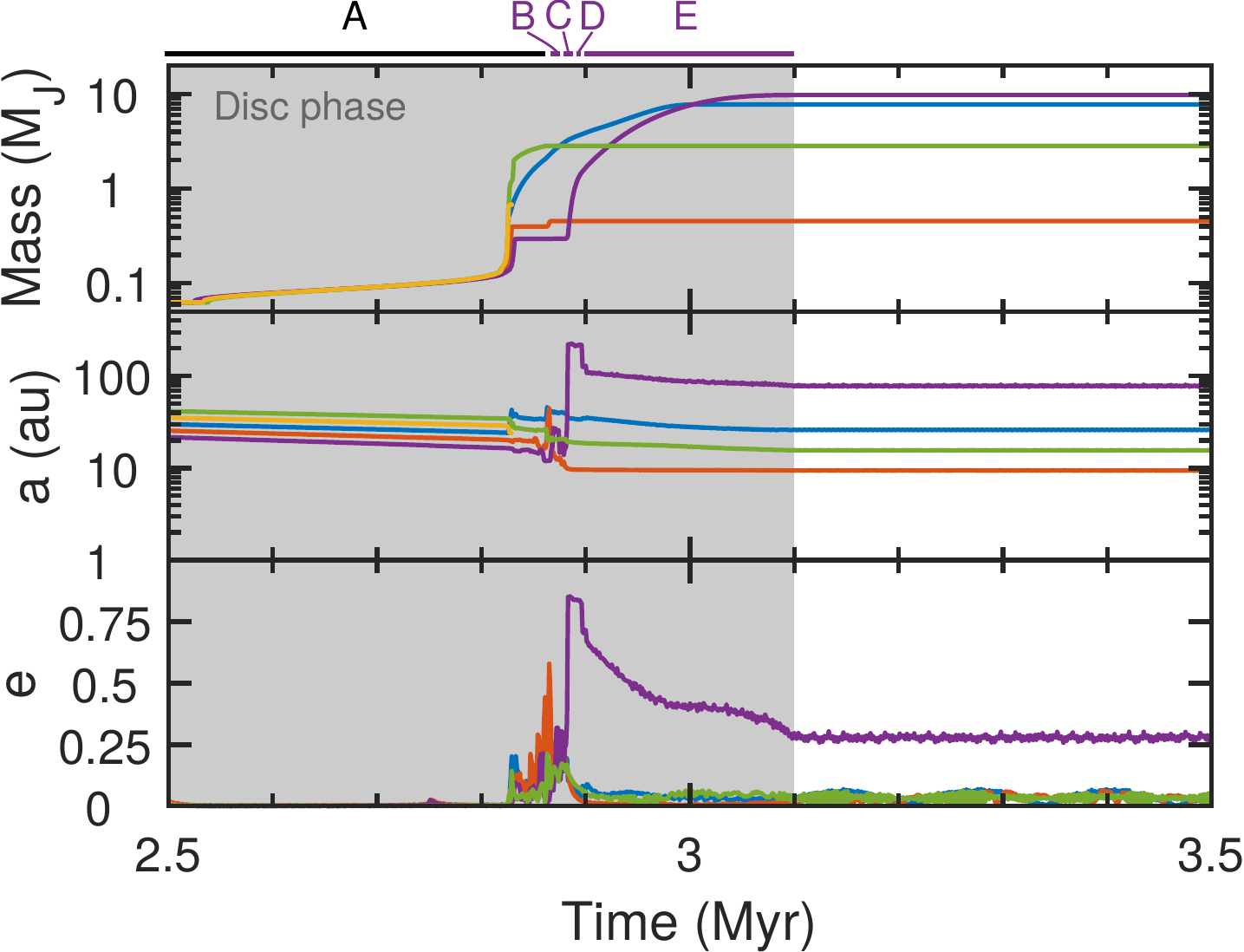}
\caption{{\it Left panel}: Evolution of planet mass against orbital distance for an example simulation.
Solid lines show the planets' semi-major axes, while dashed lines show the planets' pericentres and apocentres. 
Filled black circles represent final masses and semi-major axes for surviving planets.
Black horizontal lines show the extent of the planets' final orbits from pericentre to apocentre.
The red cross indicates the expected mass and projected orbital distance of \hdb.
{\it Right panel}: Temporal evolution of planet masses (top), semi-major axes (middle) and eccentricities (bottom).
The shaded grey area indicates the time in which the gas disc was present in the simulation. Labels in the left and right panels summarise the description in the text (Sect.~\ref{sec:examplesystem}).}
\label{fig:mva_sim}
\end{figure*}

For each simulation we placed five planets of masses $15\me \le \mp \le 20\me$ at $\rp=15$--30~au, i.e.\ in the outer disc beyond the radial structure, in close proximity to each other (initial period ratios between neighbouring planets ranging from 1.08 to 1.7).
We placed the planets in close proximity to each other to ensure that they were able to become trapped in resonant chains fairly quickly before a single core could undergo runaway gas accretion and thus destabilise the system.
This configuration is frequently seen to arise in global planet formation simulations that include planet migration, planetesimal accretion, mutual interactions between planetary embryos, and evolution of the protoplanetary disc \citep{ColemanNelson16,ColemanNelson16b}. However, due to the chaotic nature of the formation processes (i.e.\ migration, planetesimal accretion rates, $N$-body interactions), we force this initial set-up onto the planets for these simulations so as to save on computational time.
We also varied the location of the radial structure as described above and the formation time of the giant planet cores which ranged between 1.5~and 2.5~Myr.
These different initial conditions led to the computation of 792~simulations.

\begin{figure*}[th!]
\centering
\includegraphics[scale=0.6]{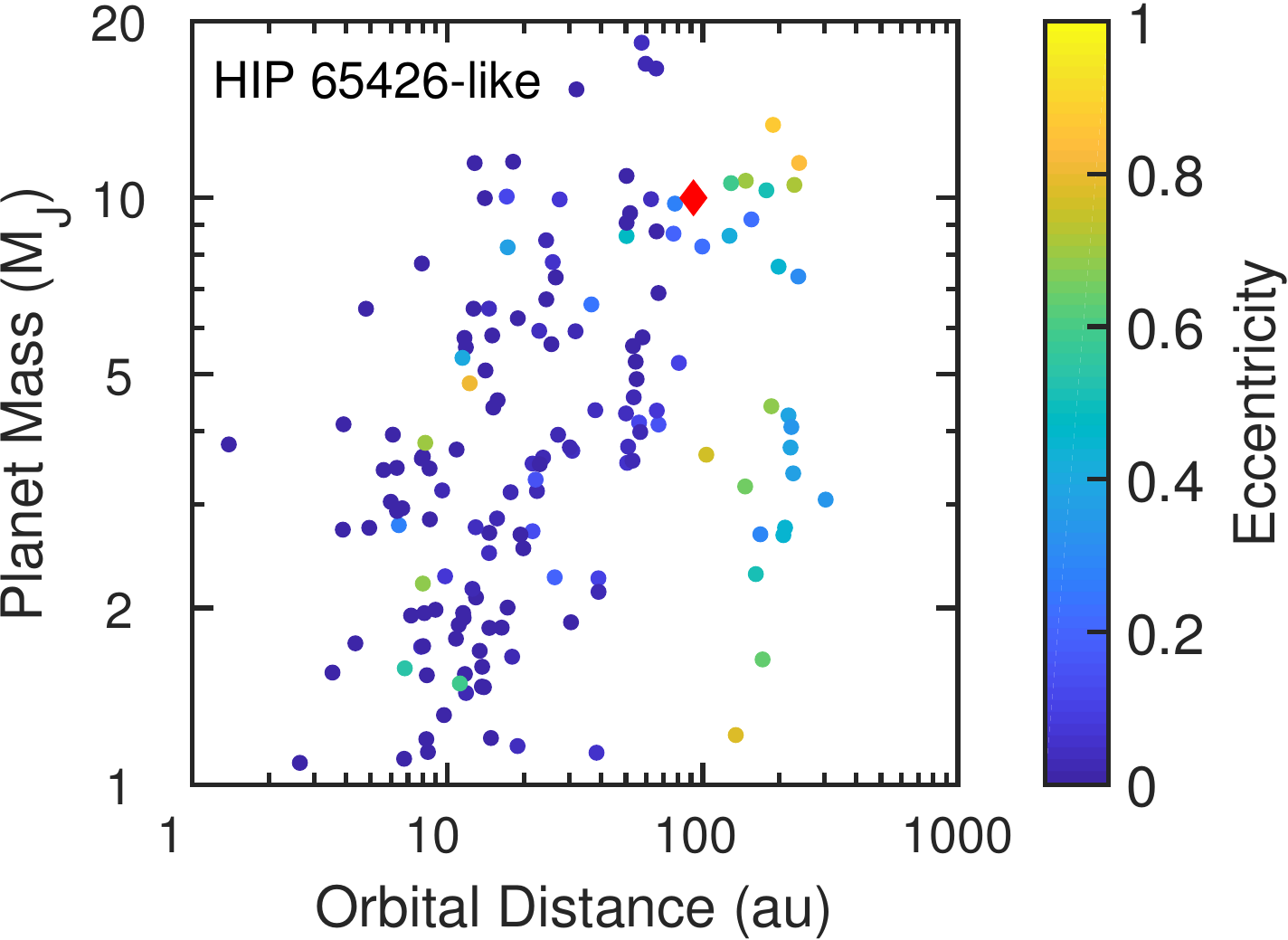}
\includegraphics[scale=0.6]{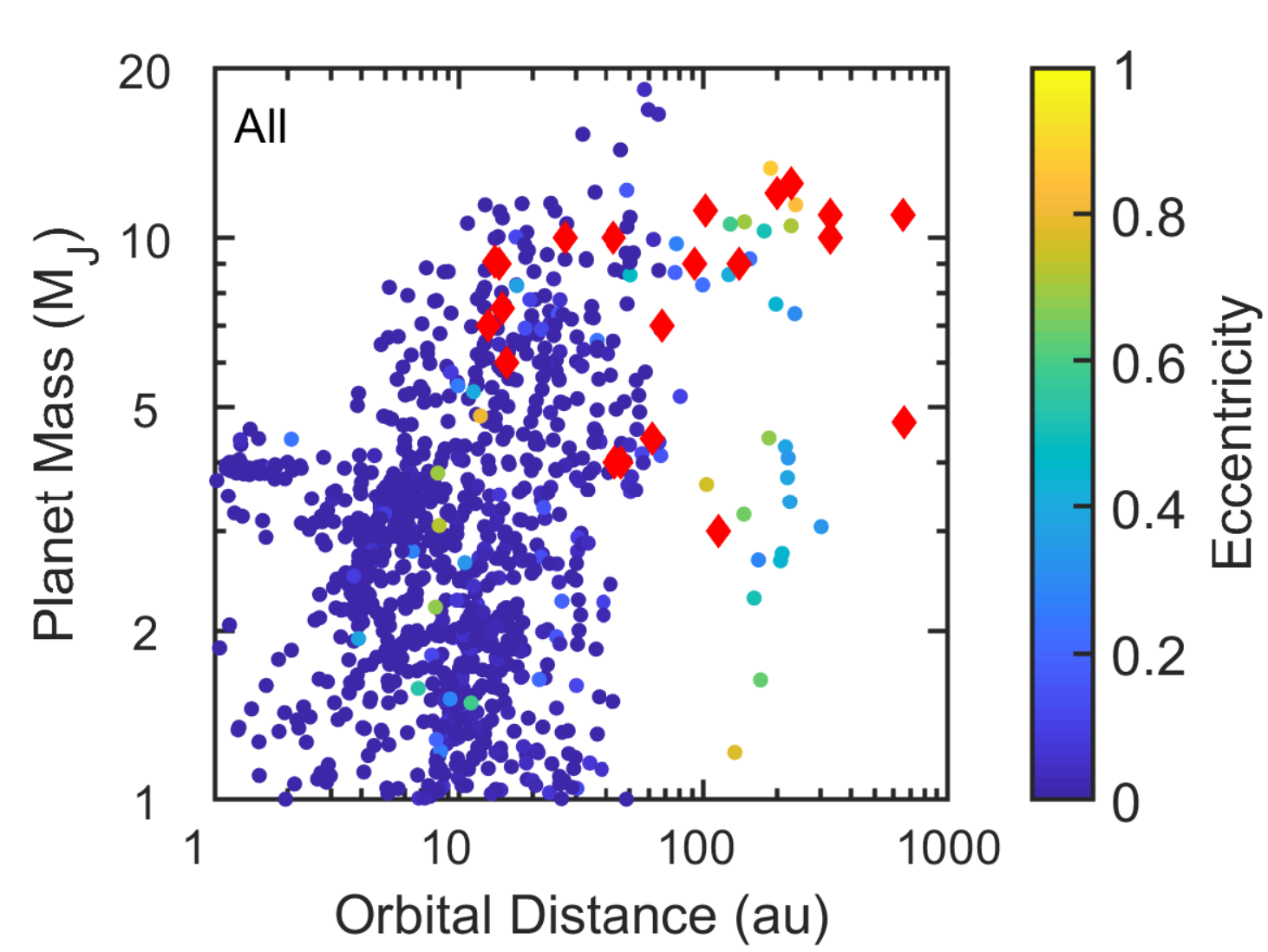}
\caption{{\it Left panel}: Final planet masses and semi-major axes from simulations that formed a giant planet with final semi-major axis $\rp>50$~au and final mass $\mp > 1~\MJ$.
Each planet's final eccentricity is colour-coded, whilst the red diamond denotes the expected mass and semi-major axis of \hdb.
{\it Right panel}: Final planet masses and semi-major axes from all simulations.
Each planet's final eccentricity is colour-coded.
Red diamonds now display the observed directly imaged planets (data taken from \url{exoplanet.eu}).
}
\label{fig:all_mva_ecc}
\end{figure*}

\subsection{Example HIP 65426-like simulated system}
\label{sec:examplesystem}

Figure~\ref{fig:mva_sim} shows the mass versus orbital distance evolution (left panel) and the temporal evolution of planet masses, semi-major axes, and eccentricities (right panel) of a typical example of such a simulation.
The mass versus orbital distance tracks of the planets are shown with solid lines indicating semi-major axes and dashed lines displaying the planets' pericentres and apocentres. Black dots represent the final masses and semi-major axes of the planets with the red cross showing the mass and orbital distance of \hdb\ as discussed in Sect.~\ref{Theil:MSi}.
As the simulation starts, all of the planets begin to accrete gas and migrate inwards towards the radial structure.
The planets' migration stalls as they approach the radial structure (see label~A in Fig.~\ref{fig:mva_sim}) due to the enhanced corotation torques arising from the radial structure's effect on the local disc profile.
The outer three cores (blue, green, and yellow lines) then undergo runaway gas accretion, opening a gap in the disc (see the sudden increase in mass for some of the planets at $\sim 2.8$~Myr in the top right panel of Fig.~\ref{fig:mva_sim}).
The inner planets (purple and orange lines) are then starved of gas due to the opening of gaps in the disc, delaying their ability to transition to runaway gas accretion.
After a further 50~kyr, the system becomes unstable with two of the giants impacting each other.
Other cores are also scattered, with one core (purple line) being scattered to $\rp \approx 220$~au with an eccentricity $\ep \approx 0.85$ (label~B in Fig.~\ref{fig:mva_sim}).
This core then undergoes runaway gas accretion at this large orbital distance, accreting gas when it enters the disc on its eccentric orbit (label~C in Fig.~\ref{fig:mva_sim}).
When the planet's mass has risen to $\mp \approx 2~\MJ$, it has a close encounter (label~D in Fig.~\ref{fig:mva_sim}) with the most massive giant in the
system (with a mass $\mp\approx4~\MJ$, shown by the blue line).
This lowers the semi-major axis and eccentricity of the former to $\rp\approx125$~au and $\ep\approx0.7$ respectively (see the drops in semi-major axis and eccentricity at $\sim 2.9$~Myr in the middle and bottom right panels of Fig.~\ref{fig:mva_sim}).
This planet then continues to accrete gas and slowly migrate in towards the central star (label~E in Fig.~\ref{fig:mva_sim}), with the disc gradually damping the planet's eccentricity.
We implement eccentricity damping for giant planets by setting the damping timescale to 100~local orbital periods.
This timescale is consistent with eccentricity damping timescales found for eccentric planets in isothermal discs \citep{Bitsch13}.

By the time the disc has fully dispersed, the planet has grown to $\mp \approx 9.8~\MJ$
and has migrated in to having a semi-major axis of $\rp \approx 77$~au.
The temporal evolution of the planets' semi-major axis, mass and eccentricity can be seen in the right panel of Fig.~\ref{fig:mva_sim}, with the shaded grey region indicating the times during which the disc is present.
Due to the circularisation of its orbit, the planet's eccentricity has dropped to $\ep\approx0.27$,
resulting in the planet orbiting between 56 and 98~au, spending most of its orbit near apocentre with an orbital distance greater than 90~au.
This range in orbital distance is shown by the horizontal black bar in Fig.~\ref{fig:mva_sim} and is compatible with the observed position of \hdb.
Also remaining in the system are three other giant planets with semi-major axes of $\rp=9.4$, 15.6, and 25.8~au, and masses $\mp=0.45$, 2.8, and 7.8~$\MJ$, respectively,
on nearly circular orbits ($\ep\lesssim0.05$ for all).

\begin{figure*}[t]
\centering
\includegraphics[scale=0.6]{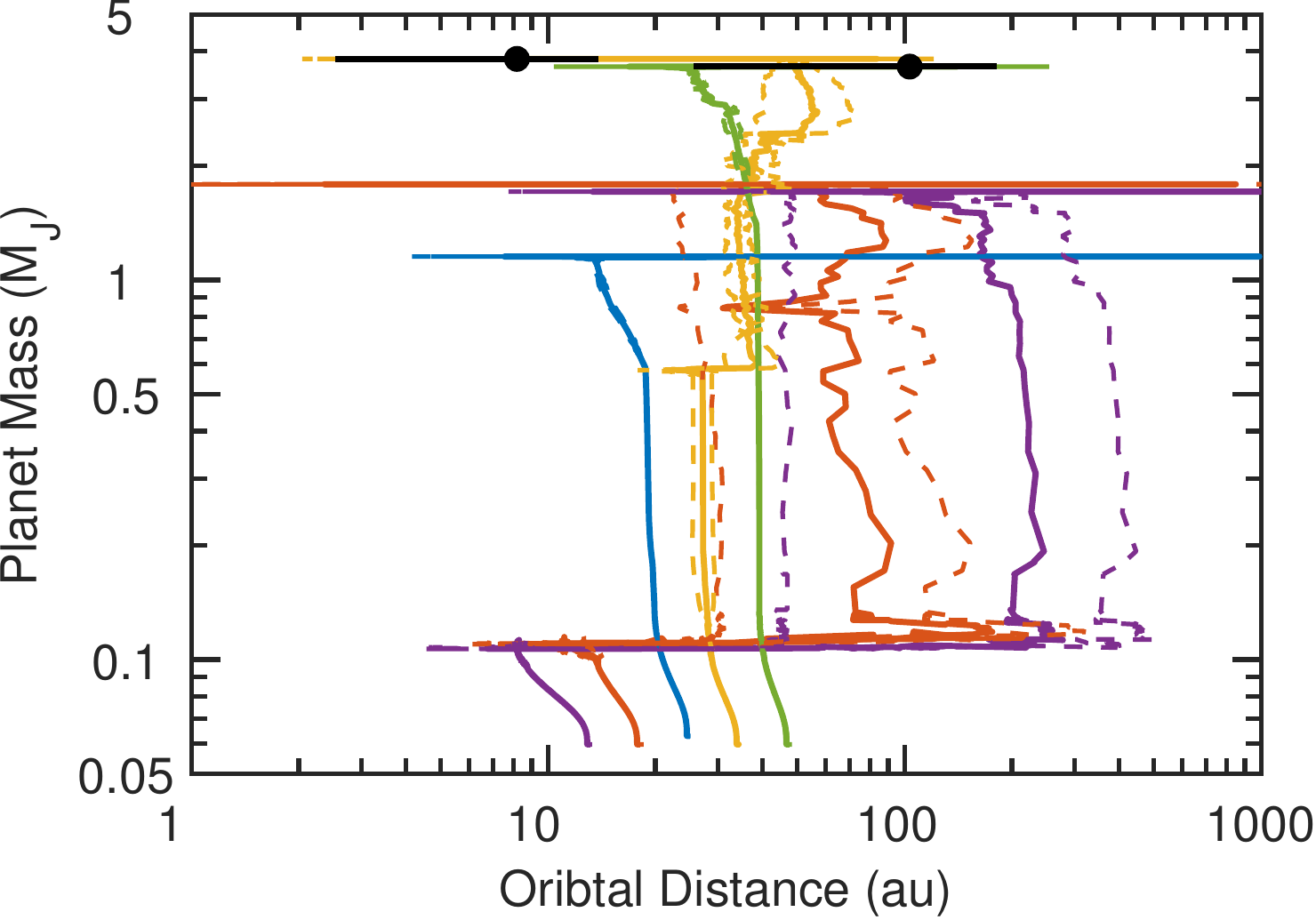}
\includegraphics[scale=0.6]{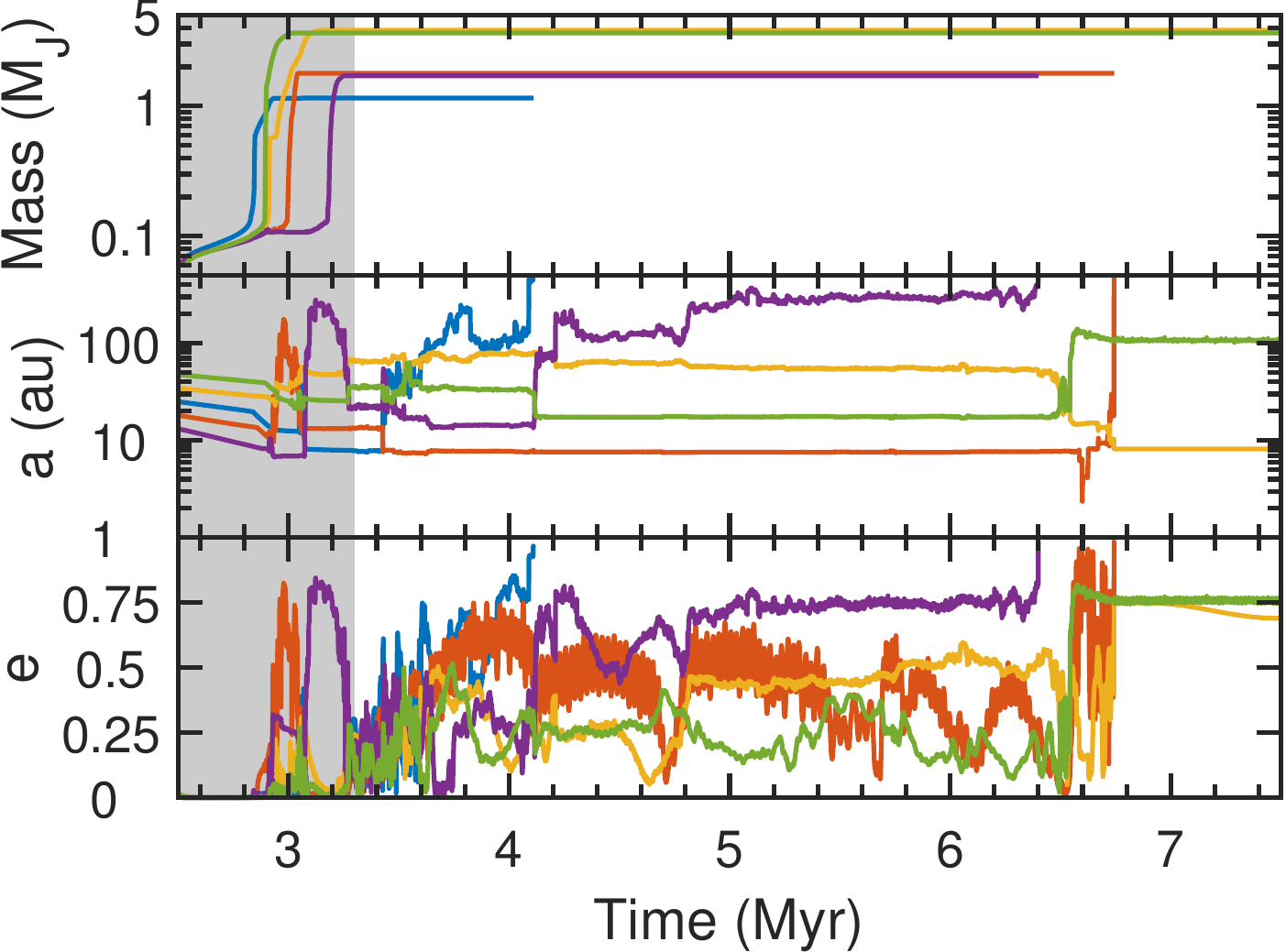}
\caption{As in Fig.~\ref{fig:mva_sim}, but for a simulation in which a giant--giant scattering event after the disc had fully dispersed (at around 4~Myr) is responsible for the final position of the \hipb-like planet.}
\label{fig:giant_giant}
\end{figure*}

\subsection{Overall results}

The system described above, with a giant planet similar to \hdb\ and numerous planets with shorter periods, is a common outcome of these simulations.
The left panel of Fig.~\ref{fig:all_mva_ecc} shows the final masses and semi-major axes from the simulations that had a similar outcome to that described above.
We define these systems as containing at least one planet with semi-major axis $\rp > 50$~au and mass $\mp > 1~\MJ$.
As can be seen, there are numerous giant planets that are similar to \hdb\ (shown by the red diamond) in terms of mass and semi-major axis (projected orbital distance for \hdb), but with a wide range of eccentricities, spanning essentially $\ep=0$--1.
All of these planets are accompanied by a number of interior giant companions that could be detectable in long-baseline radial-velocity surveys.

The right panel of Fig.~\ref{fig:all_mva_ecc} shows the final semi-major axes and masses of all surviving planets in the simulations.
The colour of the marker indicates each planet's eccentricity, and the red diamonds show the currently observed planets found in direct imaging surveys.
There is good agreement between the observations and simulated giant planets in terms of semi-major axes and planet mass.
The eccentricities cannot really be compared since
for observed planets
they are not well constrained
due to insufficient time sampling of their long orbital periods \citep{bowler18}.
The simulated giant planets have non-zero eccentricities which
in many cases are significant ($\ep>0.5$), with most of these high-eccentricity giants having semi-major axes greater than 100~au.
This is not surprising since for these planets to attain such large semi-major axes, they need to undergo significant scattering, which induces high eccentricities, and since the planets have had little time to migrate back in towards the central star, their orbits have also had insufficient time to circularise.

The plot also shows that for the distant giant planets, eccentricity and orbital distances are positively correlated, an imprint of that planet's main scattering event.
This is expected from the fact that the original formation region ($\sim20$--30~au) remains, at least for a single scattering event, part of the orbit as its pericentre distance.
In the simulations, however, eccentricity damping from the gas disc and minor interactions with other planets in the system can decrease the planet's eccentricity over time, raising the pericentre away from the formation region.
Since the distant planets that formed in the simulations had insufficient time to circularise fully, due to dispersal of the gas disc, this imprint of the main scattering event remains, explaining the distance--eccentricity correlation.

Also seen in Fig.~\ref{fig:all_mva_ecc} are numerous giant planets with semi-major axes $\rp<10$~au.
Quite often these giant planets were responsible for scattering  a giant planet core into the outer system that could then undergo runaway gas accretion at hundreds of astronomical units,
as was described in the example simulation above (Section~\ref{sec:examplesystem}).
While these planets are typically too faint and too close to the star to be observed in direct imaging surveys, they could be observed in radial velocity surveys \citep{RVSURVEY} or in astrometry surveys such as GAIA \citep{GAIA16,GAIA18}.
Further observations of \hd\ using the radial velocity or astrometry technique could yield additional giant planets in the system closer to the star than \hdb.
Very recently, \citet{cheetham19} ruled out to 5~$\sigma$ the presence of further companions more massive than 16~$\MJ$ down to orbital separations of 3~au.
This is consistent with the simulated planets shown in Fig.~\ref{fig:all_mva_ecc}, %
which all fall below the detection limits out to 30~au.
Most of the inner companions to \hdb-like planets have masses well below 10~$\MJ$.
Should these planets exist, and if \hdb\ were found to have an eccentric orbit, this could suggest  the formation origin of \hdb\ as described here, and may indicate that other directly imaged giant planets should have giant planet companions closer to their host star than has been observed up to now.

For systems that contained two or more giant planets at the end of the disc lifetime, it is possible that dynamical instabilities between giant planets as the systems age lead to the planets having wider orbits, similar to \hdb.
Figure~\ref{fig:giant_giant} shows the planet mass versus semi-major axis evolution (left panel) and the temporal evolution of planet mass, semi-major axis, and eccentricity for such a scenario.
Here the planets undergo a similar initial evolution to that described in Sect.~\ref{sec:examplesystem}, but as the disc fully disperses, all five giant planets have relatively stable orbits (given the eccentricity-damping effect of the gas) with $\rp=8$--70~au.
However these orbits are not stable on long timescales after disc dispersal, and within 0.2~Myr of the disc dispersing, the planets undergo significant dynamical instabilities, increasing eccentricities and scattering some of the giants to larger semi-major axes.
Continued interactions resulted in three of the giant planets being ejected from the system, with the two most massive giant planets remaining.
These surviving planets are shown by the black dots in the left panel of Fig.~\ref{fig:giant_giant}, where the solid black horizontal lines shows the extent of their orbit from pericentre to apocentre.
A more detailed study of forming \hdb ~through giant--giant scattering is discussed in Sect.~\ref{sec:Gps}.

\section{Post-formation scattering of giant planets}
\label{sec:Gps}
In this section we explore the formation of systems with giant planets on wide orbits, such as \hipb, through planet--planet scattering after disc dispersal. This mechanism is  known to create highly eccentric planets \citep{FoRa1996,LiId1997,Cha2008,MaWe2002}. We estimate here its efficiency in raising the apocentre of a giant planet above $\sim 100$~au without ejecting it. We examine two scenarios: two-planet scattering and three-planet scattering. A system with two planets behaves qualitatively differently than a system with three or more planets.

On initially coplanar circular orbits, if the initial semi-major axes of two planets are closer than \citep{Wi1980,DePaHo2013}
\begin{equation}
\frac{a_1-a_2}{a_1}<1.46\left( \frac{m_1+m_2}{\Mstar} \right)^{2/7},
\label{eq:deck}
\end{equation}
their orbits will be unstable on short timescales, typically of order $\tau\sim((m_1+m_2)/\Mstar)^{-1/3}P_{1}$ \citep{PeLaBo2017}, where $P_k$, $a_k$, and $m_k$ are the orbital periods, semi-major axes, and the masses of the two planets, and $\Mstar$ is the mass of the central star.
Once such a system undergoes an instability, the most probable outcome is a single-planet system
\citep{FoRa2008}. %
Being in the unstable area given by Equation~(\ref{eq:deck}) after disc dispersal requires one of two scenarios:
\begin{enumerate}
\item[(a)] the planets have migrated into a stable configuration such as a 1:2 mean-motion resonance (MMR) during the disc phase \citep{LePe2002}, and this stable configuration was disrupted after disc dispersal; 
\item[(b)] the planets were in the unstable area during their formation, but the significant disc mass
postponed the instability, for instance through eccentricity damping \citep{FoRa2008}.
\end{enumerate}
In the  case (b), instability can ensue before the total dispersal of the disc, which might still affect the orbital evolution of the giant planets. The results of this section thus have  to be compared to the occurrences of giant--giant scattering during the disc phase (Section~\ref{sec:formationscenario}).
 
On the other hand, a system with three or more planets does not have such a sharp stability boundary as in Equation~(\ref{eq:deck}). These systems can become unstable for much larger initial spacings. However, as the initial spacings increase, the timescale of the first close encounter increases as well \citep{ChaWeBo1996}. Single giant planets or pairs are common outcomes of this instability \citep{Cha2008}, as seen for example in Fig.~\ref{fig:giant_giant}.

The two- and the three-planet-scattering scenarios are both consistent with observational constraints. Out of the hundreds of giant planets of mass above $2~\MJ$ that have been observed with semi-major axes ranging from 1 to 20~au, tens are known to belong to multi-planetary systems containing at least two giant planets\footnote{See \url{exoplanet.eu}. However, these statistics being incomplete due to observational biases, the multiplicity of giant planet systems is probably underestimated.}. 
 
To explore these two scenarios, we performed $N$-body simulations of \hip-like systems after disc dispersal. We used the variable-step integrator \texttt{DOPRI}, whose behaviour for highly eccentric orbits was validated in a previous work \citep{LeRoCo2018}. We integrated the synthetic systems for $5\times 10^6$~years, which is comparable to the age of the system since disc dispersal (see Section~\ref{Theil:L-Kurven}).
Alternatively, integrations were stopped when only one planet remained in the system. The mass of the star was set to $\Mstar = 2~\Msun$, while the mass of each planet was randomly picked in the interval $m_k=5$--15~$\MJ$. The radius $R_k$ of each planet $k$ was set to $R_k=\left(1.1+0.06 \times m_k/\MJ\right)~\RJ$, which  roughly fits the non-accreting hot population of \citet{morda12_II}  %
at 3--5~Myr\footnote{This can be explored with data from the Data Analysis Centre for Exoplanets (DACE) platform at \url{https://dace.unige.ch}, in the `Evolution' section.}. Planets that entered the Roche limit of the star were removed from the simulation, with the Roche limit given by $R_{\star,\textrm{Roche}}\approx 2.2~\Rsun$ (the stellar radius is $R_{\star}=1.77~\Rsun$; \citealp{chauvin17}) for our considered range of planetary masses and radii. Collisions between planets were detected when the physical radii of the two objects intersect and were treated as completely inelastic, i.e.\ assuming perfect merging and conservation of total momentum and mass. Other collision models including possible hit-and-runs and energy dissipation might change  the outcomes of the simulations slightly. However, they should not create a significant number of broader orbits as these collisions typically reduce the eccentricities of the bodies.

Initial eccentricities were set to $\ep=0$ and will be excited though planet--planet
interactions. As we are primarily interested in the feasibility of raising the apocentre of a planet above $\sim100$~au, we restrict our study to the coplanar case and set all inclinations to $i=0$. All other angular orbital elements where chosen randomly within [0:$360^\circ$]. The initial distribution of semi-major axes depends on the considered scenario, but they are generally taken in the 10--15~au range as it is the upper limit for the typical formation of giants in the core accretion scenario as mentioned above. Taking large initial semi-major axis makes it possible, through angular momentum transfer with other planets, to raise the apocentre of a given planet to greater values without being too close to ejection.

 \subsection{Two-planet scattering}
\label{sec:2ps}

\subsubsection{Conservative case}
\label{sec:2ps konservativ}
 \begin{figure}
\begin{center}
\includegraphics[width=0.99\linewidth]{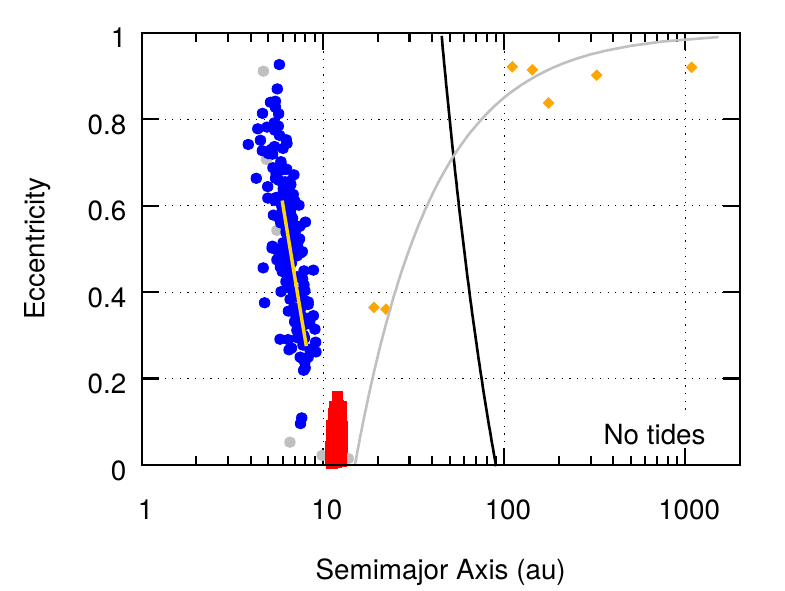}
\caption{\label{fig:scat2p_notides} Outcome of the scattering of two giant planets initially on circular orbits with semi-major axes near 10~au in the conservative (i.e.\ non-dissipative) case. The few systems that kept their two planets over $5\times 10^6$~years are displayed in grey ($\lesssim 1\,\%$ of the systems). Other systems evolved into one-planet systems typically after  $10^5$~years, either through planet--planet collision ($\approx 63\,\%$; red), ejection of the other planet ($\approx 36\,\%$; blue), or collision of the other planet with the star ($\approx 1\,\%$; orange). The yellow line is the predicted orbit of the remaining planet after an ejection for typical values (see Eq.~\ref{eq:2pEAMej}).
To the right of the black line are planets whose  apocentre is above $90$~au (projected distance of \hipb) and the grey line shows the orbits whose pericentres are at 15~au, which is the detection limit of an eventual companion of mass $\mp \gtrsim 5~\MJ$ to \hipb\ \citep{chauvin17}.}
\end{center}
\end{figure}
In the two-planet-scattering scenario, the inner planet was positioned at $\rp=10$~au, while the outer one was positioned slightly inside the instability domain (Eq.~\ref{eq:deck}), at $a_2=a_1\left(1+1.42 ((m_1+m_2)/\Mstar\right)^{2/7}$. We integrated 500~systems with this set of initial conditions; the  final outcomes are plotted in Figure~\ref{fig:scat2p_notides}.

In the conservative case, two planets on intersecting orbits will continue to experience close encounters until one of the three following outcomes happens: planet--planet collision, planet--star collision, or planet ejection. These events occurred within a few $10^5$~years, which is significantly shorter than the estimated age of the system. We now discuss each in turn.

Planet--planet collisions tend to decrease the eccentricity that the planets acquired during their stay in the unstable domain, while energy conservation ensures that the semi-major axis of the resulting planet $a_{\rm r}$ lies between the semi-major axes of the initial ones \citep{FoRa2008}: %
\begin{equation}
\label{eq:2pEcoll}
a_{\rm r}= \frac{a_1a_2(m_1+m_2)}{m_1a_2+m_2a_1}.
\end{equation}
This is the most common outcome ($\approx63\,\%$ of all systems), leading to the red clump between 10 and 15~au in Fig.~\ref{fig:scat2p_notides}. These semi-major axes are too small to correspond to \hipb.

In the case of ejection, the escaping planet typically leaves the system with a very low (positive) energy \citep{MoAd2005}. The orbit of the remaining planet is hence predictable using angular momentum and energy conservation, yielding
\begin{subequations}
\label{eq:2pEAMej}  %
\begin{align}
 a_{\rm r} &= \frac{a_1a_2 m_{\rm r}}{m_1a_2+m_2a_1}, \label{eq:2pEAMej a} \\
e_{\rm r} &=\sqrt{1-\left( \frac{m_1 \sqrt{a_1}+m_2 \sqrt{a_2}-m\ej\sqrt{2q\ej}}{m_{\rm r} \sqrt{a_{\rm r}}}\right)^2 }, \label{eq:2pEAMej b}
\end{align}
\end{subequations}
where the subscripts `r' and `ej' refer to the remaining and ejected planet, respectively, and where $q\ej$ is the ejected planet's minimal distance to the star on its parabolic orbit. This scenario represents almost all other cases ($\approx36\,\%$ of the systems). The range of possible orbits for the typical values $m_1+m_2=20~\MJ$ and $q\ej=10$~au is shown by the yellow line in Figs.~\ref{fig:scat2p_notides} and \ref{fig:scat2p_tides}. It closely matches  the distribution from the $N$-body integrations.
This scenario again leads to planets with orbital distances too small in comparison to \hipb.

The last outcome, planet--star collisions, is less likely for our range of initial conditions ($\approx1\,\%$), but yields a wider range of final configurations. To fall onto the star, a planet initially on a circular orbit at $10$~au needs to give most of its angular momentum to the other planet. Depending on the mass ratio of the planets, this might not be enough to eject the outer planet, which therefore would remain on a wider orbit. Figure~\ref{fig:scat2p_notides} shows that most of these remaining planets are consistent with the observed projected distance of \hipb.
In this set-up (two planets, no tides) \hipb\ would be on a highly eccentric orbit, and it would be the only body in the system. We  show later, when tides are included in the two-body scenario, that the outcomes explaining \hipb\ can again only contain the scattered body alone (the other was sent into the star), but also configurations with two remaining bodies, with the second companion very close to the star, circularised by tides.

\subsubsection{Effect of dynamical tides}
 \begin{figure}
\begin{center}
\includegraphics[width=\Bildbreite{0.99}\linewidth]{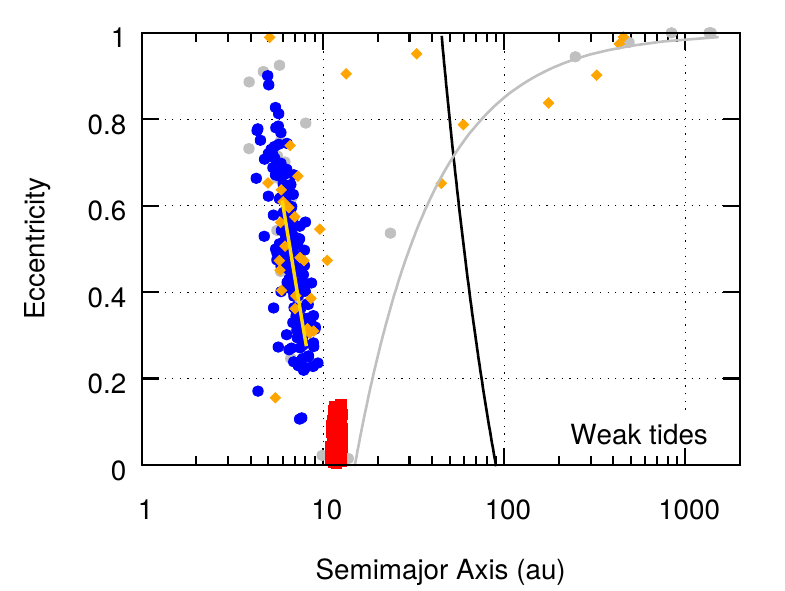}\\
\includegraphics[width=\Bildbreite{0.99}\linewidth]{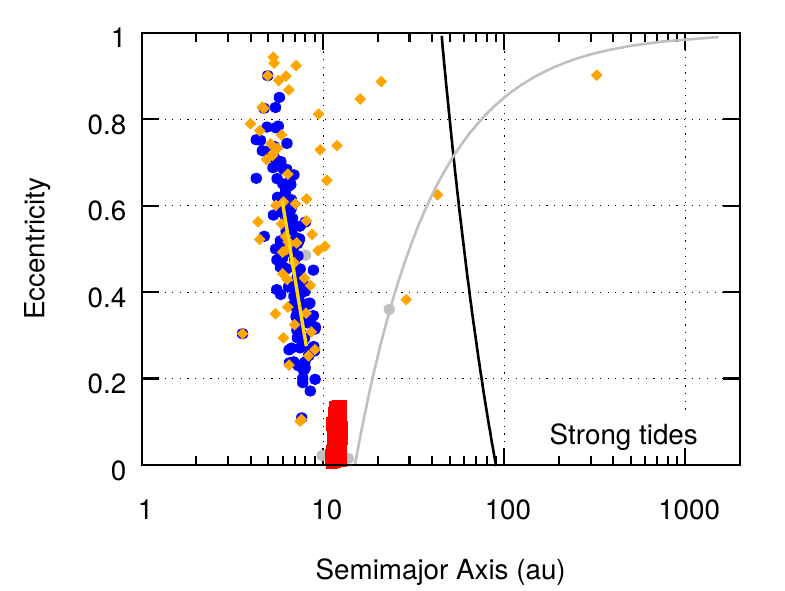}
\caption{\label{fig:scat2p_tides}%
As in Fig.~\ref{fig:scat2p_notides}, but with tidal dissipation.
\textit{Top panel:} Weak tides (Eq.~\ref{eq:Wtides}). Outcomes: planet--planet collision ($\approx 62\,\%$; red), ejection of the other planet ($\approx 33\,\%$; blue), collision of the other planet with the star ($\approx 5\,\%$; orange). The systems displayed in grey ($\approx 1\,\%$) retain two planets until the end of the simulation (5~Myr), but will eventually evolve into one of the other three configurations. 
\textit{Bottom panel:} Strong tides (Eq.~\ref{eq:Stides}).
Outcomes: planet--planet collision ($\approx 62\,\%$; red), ejection of the other planet ($\approx 25\,\%$; blue), collision of the other planet with the star ($\approx 13\,\%$; orange).}
\end{center}
\end{figure}
In the  conservative case, we saw that almost all systems with two planets initially at $\approx 10$~au underwent close encounters and ended up in the planets colliding or one of them being ejected. However, even after gas dispersal, a planet that gets close enough to the star on its orbit, typically with a pericentre $q=a(1-e)\sim R_\star \sim 0.01~\textrm{au}$,
will undergo tidal circularisation \citep{IvPa2004}. This process has the main effect of lowering the apocentre of an eccentric planet, while keeping the pericentre roughly constant, which can stop the orbits from intersecting before the ejection of the outer planet.

Since the planets that experience tidal effects will be on wide eccentric orbits, we  consider dynamical tides, which are a succession of tidal excitation (when the planet is close to its pericentre) and relaxation (during the rest of the orbit) \citep{IvPa2004}. For giant planets, the migration and eccentricity timescales of these tides can be below $10^5$ years \citep{NaIdBe2008} and hence can be comparable to the lifetime of the two-planet systems integrated in the conservative case. To take these tides into account, we adopt the formula of \cite{IvPa2004} who calculated analytically the strongest normal modes, the $l=2$ fundamental modes, of the tidal deformation. Depending on the rotation of the planet that undergoes tidal circularisation, they derived the tidally gained angular momentum ($\Delta L\tide$) and energy ($\Delta E\tide$) during a single pericentre passage in two extreme cases:

\begin{itemize}
 \item an initially non-rotating planet, which tends to maximise the effect of the tides, with
\begin{subequations}
\label{eq:Stides}
\begin{align}
\Delta L\tide & \approx - \frac{32\sqrt{2}}{15} \omega_0^2 Q^2\zeta_k \exp\left( - \frac{4\sqrt{2}}{3}\omega_0\zeta_k \right) L_k, \\
\Delta E\tide & \approx - \frac{16\sqrt{2}}{15} \omega_0^3 Q^2\zeta_k \exp\left( - \frac{4\sqrt{2}}{3}\omega_0\zeta_k \right) E_k,
 \end{align}
\end{subequations}
where $L_k=m_k(Gm_kR_k)^{1/2}$ and $E_k=Gm_k^2/R_k$ are the angular momentum and energy scales, $\zeta_k=(m_k q_k^3)^{1/2}(\Mstar R_k^3)^{-1/2}$, $\omega_0$ is the dimensionless frequency of the fundamental mode (normalised by the internal dynamical frequency $[G m_kp/R_k^3]^{1/2}$), and $Q$ is a dimensionless overlap integral that depends on the planetary interior model;

\item a planet that is spinning at the critical rotation rate, for which the passage at pericentre does not provide an increase in angular momentum. This minimises the effect of the tides:
\begin{subequations}
\label{eq:Wtides}
\begin{align}
\Delta L\tide  & = 0, \\
\Delta E\tide & \approx - \frac{1}{5\sqrt{2}} \frac{\omega_0 Q^2}{\zeta_k} \exp\left( - \frac{4\sqrt{2}}{3}\omega_0\zeta_k \right) E_k.
\end{align}
\end{subequations}
\end{itemize}
We translate either of these two expressions for $\Delta L\tide $ and $\Delta E\tide$ into migration and eccentricity damping timescales using respectively \citep{NaIdBe2008}
\begin{subequations}
\label{eq:taus}
\begin{align}
\tau_a & = -\frac{a_k}{\dot a_k}= \frac{Gm_k \Mstar}{2a_k}  \frac{P}{-\Delta E\tide}, \\
\tau_e & = -\frac{e_k}{\dot e_k}= Gm_k \Mstar P \left( -a_k \gamma_k  \Delta E\tide + \sqrt{\frac{\gamma_k G \Mstar}{a_k {e_k}^2}} \Delta L\tide\right)^{-1},
\end{align}
\end{subequations}
with $\gamma_k=(1-e_k^2)/{e_k}^2$ and where $P$ is the orbital period of the planet.

We note that for these tidal models to be realistic, it is necessary that the normal modes arising near the pericentre passage be fully dissipated before the next pericentre passage, which is typically the case for a semi-major axis above a few astronomical units \citep{IvPa2004}. Moreover, the actual spin of the planet evolves over time, which causes the effectiveness of the tides to vary. As a result, we assume that this model allows us to correctly represent the evolution of the orbit of an inner planet during the early stages of its apocentre lowering, but does not represent correctly the final state of the inner planet. This is however not of concern as we are primarily interested in the final orbit of the outer planet.

To estimate the effect of the tides on the systems that were integrated in the conservative case, we re-ran the same set of initial conditions as in Fig.~\ref{fig:scat2p_notides} in two cases: for weak tides, using the set of equations (\ref{eq:Wtides}), and for strong tides, using the set of equations (\ref{eq:Stides}). The results are presented in the top and bottom panel of Fig.~\ref{fig:scat2p_tides}, respectively. We assumed that $\omega_0=1.2$ and $Q=0.56$ for all planets, as these dimensionless parameters tend to be independent of the radius of the planet for $m_k=5~\MJ$, and we assume that it remains the case for more massive planets \citep{IvPa2004}.

In both cases, the majority of the systems still evolve towards the two main outcomes of the conservative case: either ejection of one of the planets or planet--planet collision, leaving a single planet with a semi-major axis below 15~au. The relative occurrences are very similar to the conservative case. The similarity with the conservative case is easily understandable as the tides affect systems for which the pericentre of one of the planets goes below a few hundredths of an astronomical unit, which is relevant only for a few systems. Nevertheless, including tides increases the number of systems that exhibit a planet--star collision or that retain two planets until the end of the simulation.
The `weak tides' model produced more planets with large semi-major axes than did   the `strong tides' model. The reason is that the latter tends to lower the apocentre of the inner planet before it can exchange enough angular momentum with the outer planet. %
It is important to note that the number of planet--star collisions is considerably overestimated due to our continuous application of dynamical tides even when the apocentre of the inner planet goes below a few astronomical units. In that sense, most of these systems are more likely to retain a close-in circularised giant planet in addition to the outer ones displayed in orange in Fig.~\ref{fig:scat2p_tides}. Although the tides allow for a broader diversity of outcomes for the scattering of two giant planets, only a small fraction of the systems contain planets with apocentres above 90~au.

\subsection{Three-planet scattering}
\label{sec:3ps}
\begin{figure}
\begin{center}
\includegraphics[width=\Bildbreite{0.99}\linewidth]{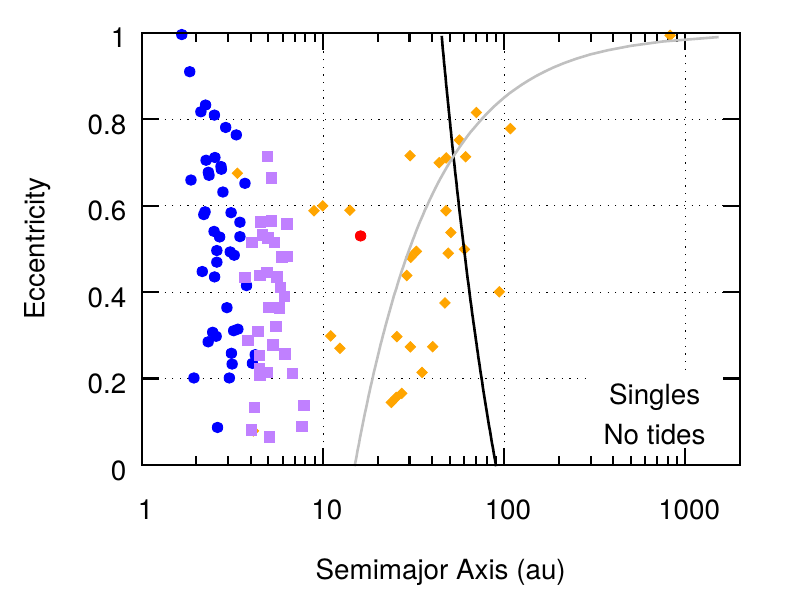}\\
\includegraphics[width=\Bildbreite{0.99}\linewidth]{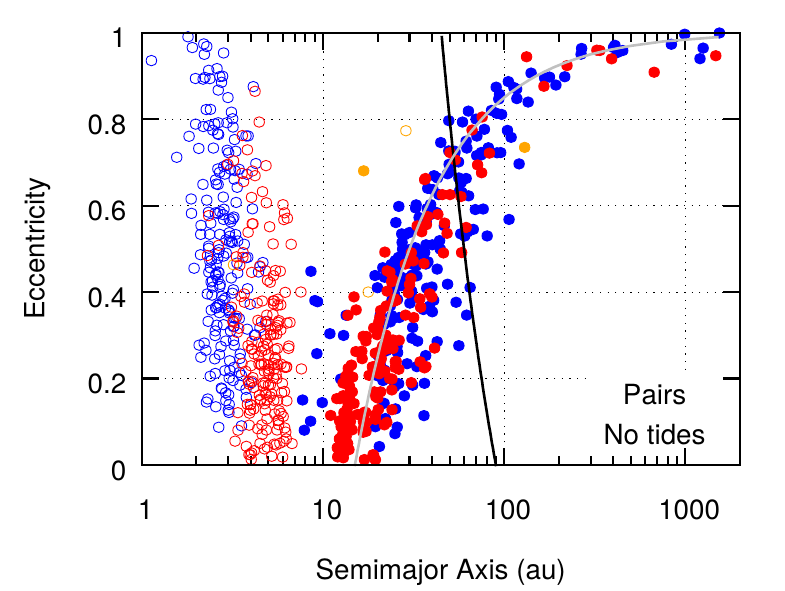}\\
\caption{\label{fig:scat3p_notides} Outcome of the scattering of three giant planets initially on circular orbits with semi-major axes near 10~au in the conservative (tide-free) case. Only the systems that lost at least one planet ($\approx46\,\%$ of the initial 1000) are represented.
The black and grey lines are as in Figs.~\ref{fig:scat2p_notides} and~\ref{fig:scat2p_tides}.
\textit{Top panel:} Systems that ended up with a single planet at the end of the run ($\approx23\%$ of the systems that underwent a strong instability). Blue dots represent planets whose two companions were ejected, purple for the systems that underwent both planet--planet collision and ejection, orange %
for those that underwent both ejection and collision with the star, and red when two planet--planet collisions occurred. 
\textit{Bottom panel:} Systems that ended up with two planets at the end of the run ($\approx77\%$  of the systems that underwent a strong instability). The colour-coding is the same as in Figs.~\ref{fig:scat2p_notides} and~\ref{fig:scat2p_tides}, with open circles for the inner planet and filled circles for the outer one.
}
\end{center}
\end{figure}

\begin{figure}
\begin{center}
\includegraphics[width=\Bildbreite{0.99}\linewidth]{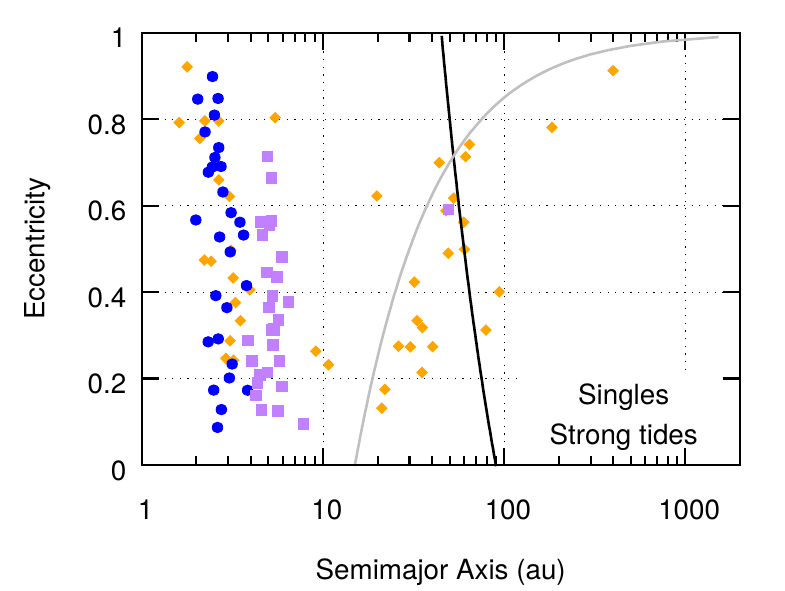}\\
\includegraphics[width=\Bildbreite{0.99}\linewidth]{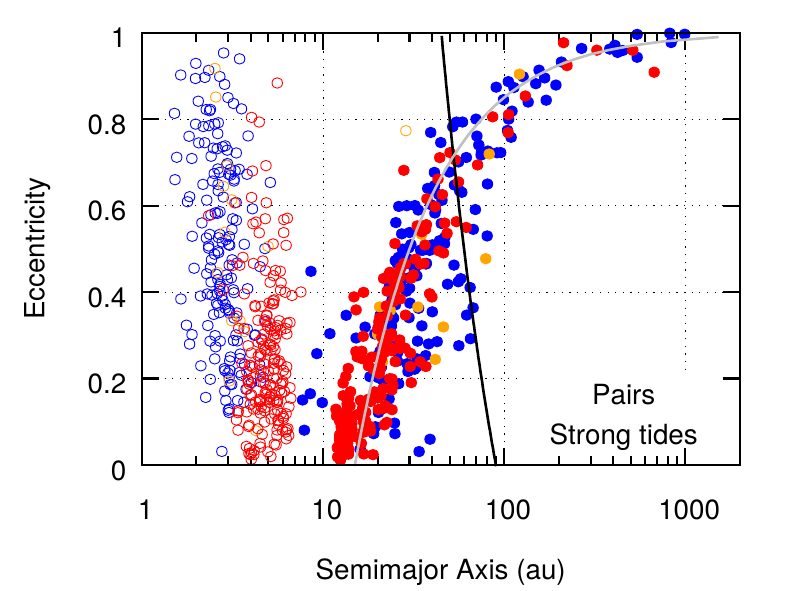}\\
\caption{\label{fig:scat3p_tides}
As in Fig.~\ref{fig:scat3p_notides}, but with strong tidal dissipation (Eq.~\ref{eq:Stides}). The top and bottom panel includes respectively $\approx 27\,\%$ and $\approx 73\,\%$ of systems that underwent strong instability.
}
\end{center}
\end{figure}

As mentioned previously, in the three-planet case there is no sharp stability condition regarding the initial semi-major axes of the giant planets. Following \cite{MaWe2002}, we initially position $m_2$ at 10~au, and $m_1$ and $m_3$ at four mutual Hill radii inside and outside the orbit of $m_2$, respectively (this corresponds to spacings $\approx5$--7~au). This initial spacing does not necessarily ensure instability in the system within the $5\times 10^6$~years of integration, and on the other hand these systems can become unstable even for much wider initial spacings, but the timescale of first encounter will increase as well \citep{ChaWeBo1996}. We chose this spacing in order to have a good probability of close encounters within the age of the system (see next paragraph).
We point out that our results are probably more general than our restricted set of initial conditions might suggest, since the time until instability for a particular set of initial conditions does not affect the statistical properties of final outcomes in this kind of study \citep{Cha2008}.

Out of 1000 initial conditions, the planets strongly interacted in $46\,\%$ of the systems, which resulted in the loss of at least one planet within $5\times 10^6$~years. These systems are shown in Fig.~\ref{fig:scat3p_notides}. In the remaining systems, the planets oscillated around their initial semi-major axis without significant increase of eccentricity and will not be discussed further. Out of the systems that interacted, we separated those resulting in single-planet systems (top panel) and two-planet systems (bottom panel). 

Single-planet systems generally underwent two ejections, a planet--planet collision and an ejection, or a planet--star collision and an ejection. Although the outcomes are less predictable than in the two-planet case, it is still planet--star collisions that tend to allow a single planet to remain on a wide orbit after the removal of its companions.

Systems that lost only one planet (bottom panel of Fig.~\ref{fig:scat3p_notides}) end up with two planets on well separated orbits, generally after an ejection or a planet--planet collision. As the eccentricity of these orbits is significant, the stability criterion used previously (Eq.~\ref{eq:deck}) is not valid. Instead, we check if the system are angular momentum deficit (AMD) stable \citep{LaPe2017,PeLaBo2017}. For coplanar orbits, the AMD of a two-planet system is given by
\begin{align}
\label{eq:AMD}
\textrm{AMD}=\sqrt{G \Mstar}\Bigg( & m_1\sqrt{a_1}\left(1-\sqrt{1-{e_1}^2}\right) \notag\\
                                 &+m_2\sqrt{a_2}\left(1-\sqrt{1-{e_2}^2}\right) \Bigg).  %
\end{align}
A given system is AMD stable if the orbits of the two planets cannot intersect through free exchange of AMD between the two planets \citep{LaPe2017}. This criterion is valid as long as the two planets are not in mean-motion resonance. For completeness, we also check if the systems are in the chaotic area due to the overlap of first-order MMRs, which is given by \citep{PeLaBo2017}
\begin{equation}
\frac{a_{\rm int}}{a_{\rm out}} < 1-1.36\left(\frac{m_1+m_2}{\Mstar}\right)^{1/5} c_{\rm min}^{1/10},
\label{eq:overlap MMR}
\end{equation}
where
\begin{equation}
 c_{\rm min} < 2 \left(2- \sqrt{1-e_1^2} -\sqrt{1-e_2^2}\right).
\end{equation}
For our considered range of masses, this criterion (Eq.~\ref{eq:overlap MMR}) is valid when both eccentricities $e_k\gtrsim 0.2$. We find that more than 99\,\% of the resulting systems with two planets are AMD stable. 

The two-planet systems represented in Fig.~\ref{fig:scat3p_notides} are significantly more diverse than in the two-planet scattering case (cf.\ Fig.~\ref{fig:scat2p_notides}). They generally have an inner planet with a semi-major axis comparable to or lower than the initial innermost planet, while the outer planets (filled circles) have their pericentre distributed around 15~au (grey curve). This means that, roughly, their pericentre remains near their initial semi-major axis. However, the departure from this curve can be significant.

For comparison, we re-ran the same initial conditions with strong tides (Equations~\ref{eq:Stides}). The result is displayed in Figure~\ref{fig:scat3p_tides}. The effect on the final $\ep$--$\rp$ distribution is clearly less important than in the two-planet scattering case (cf.\ Figs.~\ref{fig:scat2p_notides} and~\ref{fig:scat2p_tides}), which implies that the ejections or planet--planet collisions tend to occur before the pericentre of the innermost planet reaches a few hundredths of an astronomical unit. In fact, only approximately 1\%\  of the inner planets see their pericentre drop below 0.1~au throughout their orbital evolution.

In total, a significant fraction of the systems ends up with a planet on a wide orbit, with an apocentre several times higher than the initial semi-major axis. If we compare this to the projected distance of \hipb\ (92~au), with our choice of initial conditions $\approx 18\,\%$ of the systems ended up with a planet whose apocentre is above 90~au after $5\times 10^6$~years in the conservative case ($7\,\%$ of the single planets and $21\,\%$ of the two-planet systems), against  $\approx 16\,\%$ when strong tides are modelled.

\subsection{Conclusion about giant planet scattering}

\begin{figure}
\begin{center}
\includegraphics[width=\Bildbreite{0.99}\linewidth]{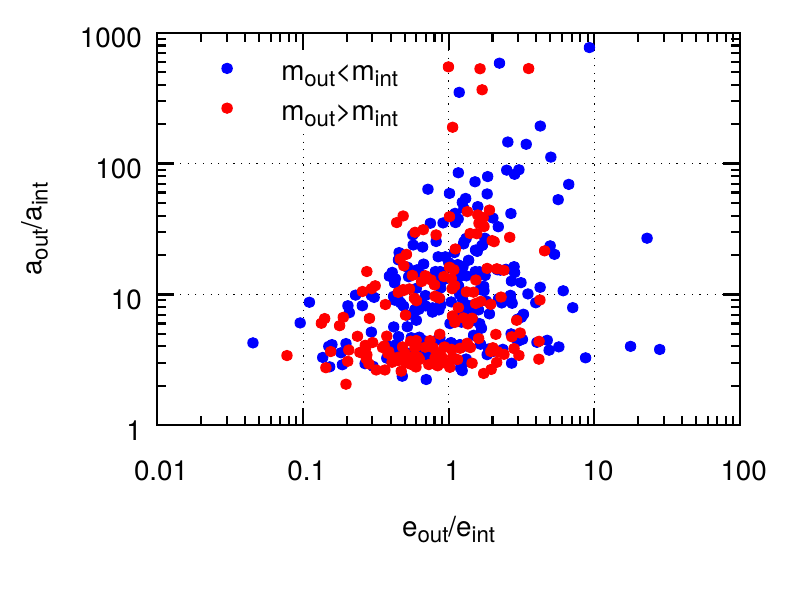}\\
\includegraphics[width=\Bildbreite{0.98}\linewidth]{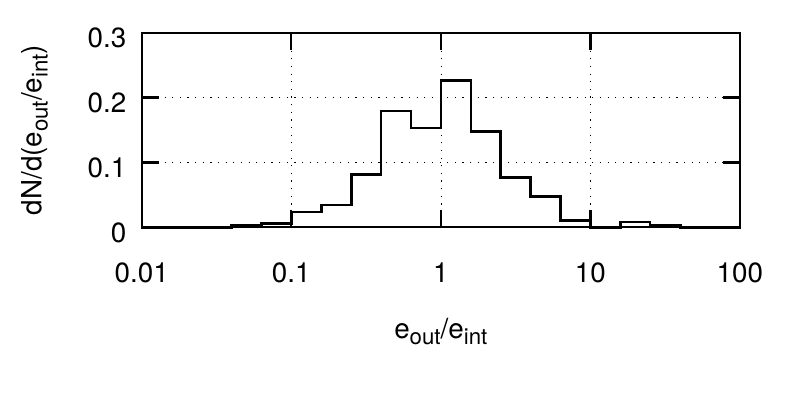}  %
\caption{\label{fig:inner} Comparison of the orbits of the inner and outer planets of each system represented in Figure \ref{fig:scat3p_notides}. In the top panel, the colour indicates which planet is the more massive: the inner one (red dots) or the outer one (blue dots). Axes show the ratio of the semi-major axes and of the eccentricity, respectively. The bottom panel displays the distribution of $e_{\textrm{out}}/e_{\textrm{int}}$. We note the logarithmic horizontal axis.}
\end{center}
\end{figure}

Both two-planet and three-planet scattering scenarios are able to create systems with giant planets on wide orbits, with
semi-major axes
above 100~au, even starting with planets in the vicinity of 10~au. In both cases, these planets tend to be highly eccentric ($\ep\gtrsim 0.5$, and generally more) as they retain a pericentre close to their initial semi-major axis. However, the occurrence rate of these orbits, as well as the presence and properties of an eventual giant planet companion, greatly depend on the studied scenario:
\begin{itemize}
\item In two-planet scattering, we have found that at most a small fraction of the systems (depending on the chosen tidal model) end up with a planet on a stable orbit with an apocentre significantly raised with respect to their initial semi-major axis. However, the instability between two planets may occur while at least a partial disc is remaining, which may lead to a broader range of outcomes (see Sect.~\ref{sec:formationscenario}). Although most of the systems that
ended up with a planet on a wide orbit (with an apocentre significantly larger than its initial one) were single-planet
systems, a proper model of the tides can circularise an inner planet on a tight orbit instead of letting it migrate all the way into the star. This would cause \hip\ to have another giant planet on a much shorter orbital period, possibly observable using the radial-velocity method.
However, it would  not be observable with current direct-imaging techniques 
such as Sparse Aperture Masking,
which push down to a few~au for this system \citep{cheetham19}.

\item In the three-planet case, the outcomes are much more diverse. In $\sim 3/4$ of the cases, two planets remain in the system on stable orbits. Most of these systems  have a planet with a semi-major axis significantly higher than initially and an inner planet with a semi-major axis comparable to the initial one or lower (see Figs.~\ref{fig:scat3p_notides} and~\ref{fig:scat3p_tides}). Figure~\ref{fig:inner} shows the $\rp$ and $\ep$ ratios between the inner and outer planets of Fig.~\ref{fig:scat3p_notides}, and which planet of the pair is more massive. There is no tendency for either the inner or the outer planet to be the more massive one, and both planets tend to have comparable eccentricities (histogram in Fig.~\ref{fig:inner}, bottom panel). If directly imaged planets such as \hipb\ obtained their wide orbit through planet--planet scattering of three giant planets initially in the 3--20~au range, it is probable that these systems also contain an eccentric inner planet with a semi-major axis of or greater than a few astronomical units. Observations are not yet constraining enough to confirm or refute the existence of such a planet around \hip\, as the current limits only exclude a planet more massive than $5~\MJ$ outside of 15~au \citep{chauvin17}, or a planet more massive than $16~\MJ$ outside of 3~au \citep{cheetham19}, and we recall that in the three-planet scattering case the remaining inner planet would not necessarily be more massive than \hipb\ (see Fig.~\ref{fig:inner}).

\end{itemize}

\section{Conclusion}
 \label{Theil:Zus}

\subsection{Summary}
The planet imager SPHERE \citep{beuzit08,beuzit19} recently revealed a companion to the 2~$\Msun$, $14\pm4$~Myr Lower Centaurus-Crux
group member \hip.
The initial analysis by \citet{chauvin17}
showed it to be of planetary mass, with $\mp=6$--12~$\MJ$, while located at a separation from its host star
(projected: 92~au) at which formation by core accretion is not expected to be efficient.
Combined with the star's unusually high rotation rate ($\vstar\approx300$~km\,s$^{-1}$),
this motivated us to take a closer look at the system to
(i)~infer joint constraints on the mass and initial (post-formation) entropy,
(ii)~explore the formation of wide-orbit (directly imaged) planets by core accretion,
and (iii)~derive predictions about the presence of further companions in the system.
While we focused on \hipb, it is an excellent representative for the relatively
recent and modestly populated class of directly imaged exoplanets in terms of mass, age, and separation from its host star.

First, we derived constraints on the mass and initial entropy of \hipb\ from its age
and luminosity (Section~\ref{Theil:MSi}).
Assuming it formed by core accretion (CA), we argued that \hipb\ should be
roughly $\DtForm\approx2$~Myr younger than its host.
We considered different priors on the mass
and entropy, including for the first time 
the mass and post-formation entropy distribution of the \citet{3M} population synthesis.
The simple but robust 2D fits for $\d^2 N/(\d \mp\,\d\spf)$
in Eqs.~(\ref{Gl:Prior beide})--(\ref{Gl:MPrior}) may be useful in other work.
Flat priors yielded $\mp=9.8  ^{+ 1.5 }_{- 2.0 }~\MJ$, whereas
the priors from the hot and cold population from the population synthesis
lead to
$\mp=9.9^{+1.1}_{-1.8}~\MJ$  %
and
$\mp=10.9^{+ 1.4}_{- 2.0}~\MJ$,  %
respectively.
Independently of the priors, the minimal post-formation entropy could be constrained
to $\spf\gtrsim9.2~\kB\,\textrm{baryon}^{-1}$.
Using the population synthesis priors made a large difference,
providing an upper bound and yielding
$\spf=10.4  ^{+ 0.7 }_{- 0.2 }$ %
in the hot-population case and
$\spf=10.2  ^{+ 0.3 }_{- 0.7 }$  %
for the cold one.

Next, we studied the formation of wide-orbit gas giants by core accretion (Sect.~\ref{sec:formationscenario}).
The idea is to let a core that formed
in the inner disc be scattered by a companion into the outer disc,
where it can undergo runaway accretion. If this scattering
happens late enough, the finite lifetime of the disc
combined with the slower type~II migration rate should
allow the planet to stay at large semi-major axes.
To counter the fast type I migration while the core forms,
we included, as in \citet{ColemanNelson16b}, a specific radial structure
which acts as a planet trap and could be due to zonal flows.

This scenario was seen to work well, producing \hipb-like planets in a
number of cases (Fig.~\ref{fig:all_mva_ecc}). In almost all systems, they were 
accompanied by interior giant companions that could be detectable  %
in long-baseline radial-velocity surveys.
Another possibility is of instabilities after disc dispersal. This too was
shown to be a possible origin for \hip-like systems, again with the prediction
of further interior companions

Finally, we focused on the post-disc phase with $N$-body integrations
of two- or three-planet systems including tides (Section~\ref{sec:Gps}).
Systems with two planets usually ($\sim2/3$ of the time) featured a planet--planet collision,
with almost all other cases ending up with a planet ejection.
For both outcomes, the remaining planet still retained too small a semi-major axis ($\rp\sim10$~au)
to explain \hipb.
In the case of three planets initially, roughly half of the systems did not
interact significantly within 5~Myr.
Of the others, about~1/4 lost two planets, with the remaining planet
matching \hipb\ only a small fraction of the time.
Systems with two remaining planets however had more diverse configurations
in the $\rp$--$\ep$ plane.
For our choice of initial conditions, $\sim1/5$ of the systems ended up
with a planet with an apocentre above 90~au (\hipb's projected separation).
We also looked at the effect of tidal circularisation, which 
can affect the orbit of highly eccentric planets that pass close to the star.
We showed, however, that in both the two- and three-planet scenarios
the outcomes are changed only slightly (Figs.~\ref{fig:scat2p_tides}
and~\ref{fig:scat3p_tides}).

\subsection{Discussion}

The main implications of our study are the following:
\begin{enumerate}
 \item We estimate a mass of 
       $\mp=9.9^{+1.1}_{-1.8}~\MJ$  %
       using the hot population and
       $\mp=10.9^{+ 1.4}_{- 2.0}~\MJ$  %
       with the cold-nominal population
       for \hipb.  %

 \item As for almost all other directly imaged planets, we find that \hipb\ is
       not consistent with the extreme cold starts \`a la \citet{marley07}.
       This is also in agreement with recent theoretical work \citep{berardo17,mkkm17}.

 \item A more precise mass determination is hindered here less
       by systematics between the different atmospheric models (see Fig.~\ref{Abb:c21-Kurven})
       than by the large relative uncertainty on the stellar age.
       The uncertain formation time $\DtForm$ is subdominant to this.

 \item Both runaway accretion at a large separation after outward scattering
       of the core as well as post-disc-phase scattering of inner gas giants
       were seen as viable scenarios to explain \hipb-like objects.

\item Our simulations show that if it formed through core accretion, \hipb\ likely has some eccentricity.
This eccentricity arises from scattering with other planets in the system.
If these scattering events occur before the end of the disc lifetime, damping with the gas disc can act to reduce the eccentricity.
In this case the planet would have a modest eccentricity, $0\le\ep\le0.5$, where the time of scattering with respect to the end of the disc lifetime determines how much eccentricity can be damped.
If the scattering event took place after the end of the disc lifetime, then the eccentricity can be higher, depending on the scattering conditions.
Therefore, if future observations revealed the eccentricity to be $\ep<0.5$,
this would not rule out the scenario of scattering before disc dispersal.
It would, however, make scattering at the end of or after the disc phase unlikely,
unless we could invoke another kind of eccentricity-damping mechanism.

The high-eccentricity cases are in contrast with the very tentative result that directly imaged planets might tend to have low eccentricities.
However, this is mostly  based on  a relatively small number of upper limits \citep{bowler18}, and the few cases with more robust determinations
are not likely candidates for the scenario presented here.

For example, several authors have favoured $\ep\lesssim0.2$ for $\beta$~Pic~b
\citep{wang16,lagrange18}, which might suggest it did not form by the scenario shown here.
While \citet{dupuy19} recently excluded $\ep<0.1$ at $>2\,\sigma$,
their derived eccentricity was only $\ep=0.24\pm0.06$.
Independently of the (modest) eccentricity of $\beta$~Pic~b, however,
its low semi-major axis $\rp\approx12$~au makes it a somewhat unlikely candidate
for formation by scattering. Also, the presence of a debris disc
makes any speculation about its dynamical origin more challenging.

As for the specific case of the HR~8799 planets, it is unlikely that
a scenario such as that studied here is responsible for their formation,
irrespective of their exact eccentricities (which appear to be low
to moderate; \citealp{wang18}).
Indeed, this would require an unlikely series of interactions without
ejections, for example four times in a row.

In any case, a longer coverage of the orbits will be necessary
to improve the statistics of the eccentricity determinations.

 \item If directly imaged planets such as \hipb\ obtained their wide orbit through planet--planet scattering of three giant planets initially in the 3--20~au range, it is probable that these systems also contain an eccentric inner planet with a semi-major axis equal to  or greater than a few astronomical units.
 
 Some previous studies have put upper limits to the existence of inner companions
 in systems with wide-orbit planetary-mass companions
 (see \citealp{bryan16} and references therein). However, these limits
 typically reach down to several tens of astronomical units for several Jupiter masses,
 and thus
  leave open a parameter space consistent with our tentative prediction. This
 could be tested by future radial-velocity surveys or further direct-imaging observations.
\end{enumerate}

We can wonder whether the inferred initial entropy %
will reveal clues  to the location of the runaway gas accretion phase.
In the absence of detailed studies of this question, the answer seems complex since
several effects are relevant at the same time:
\begin{itemize}
 \item For a given planet mass, radius (or entropy), and accretion rate,
  it is easy to show that the properties of the accretion shock onto the planet
  will not depend, at least not directly, on its location in the disc (\citealp{mkkm17},
  Marleau et al., subm.).
  
 \item The Core Mass Effect \citep{morda13} predicts higher post-formation entropies for
  higher planetesimal surface density, as should be found closer in to the star.
  
 \item Berardo et al. (2017) showed that the important quantity determining
the influence of the shock is the pre-runaway entropy of the protoplanet.
This in turn might be different for planets formed at different locations,
but the magnitude of the effect is challenging to assess without dedicated simulations.
\end{itemize}
Further  factors might come into play, such as the metallicity of the gas.
Depending on which way the different effects go, the location of runaway gas accretion
may or may not be imprinted in the post-formation entropy.
Clearly, a global dedicated study is warranted here.

Thanks to the large separation and super-Jupiter mass of the companion,
\hip\ represents an important system to explore the dynamical interactions of (proto)planets and the limits of planet formation by core accretion.
Recent studies have reached contrasting results about gravitational instability (GI),
arguing that it must be an intrinsically rare process \citep{forgan13,vigan17}
or rather that it is common but associated with very fast migration of the clumps
(see \citealp{nayakshin17} and discussion therein as well as \citealp{vorobyov18}).  %
However, these studies looked mainly at FGK stars, whereas the planet occurrence rate
seems to increase with stellar mass \citep{bowler16}.
In any case, it would be interesting also to perform a  study similar to the present one
in the context of GI, following self-consistently the formation of the central star
(see e.g.\ \citealp{nixon18}) and trying to explain its high spin frequency.
Also, predictions of the post-formation entropies and luminosities
in GI formation models (e.g.\ \citealp{forgan13,forgan15}) would be
a welcome counterpart to those of core accretion \citep{3M}.
With an orbital period $P\approx 400$--2000~years \citep{cheetham19}, approximately five to ten years are required until
the eccentricity of \hipb\ can be robustly determined if orbital curvature begins to be resolved
(G.\ Chauvin 2019, priv.\ comm.).
However, this -- and more generally for other systems too -- will be an important constraint on the formation model. Similarly, further
radial-velocity monitoring of the host to reveal or rule out
the presence of further companions would also help confirm or exclude
some of the formation pathways discussed here.

\begin{acknowledgements}
We thank G.~Chauvin for helpful comments about the manuscript
and answers to observational questions, A.-M.\ Lagrange for the interesting discussions,
and the referee for the comments and questions that helped clarify the manuscript.
G.-D.M.\ and C.M.\ acknowledge support from the Swiss National Science Foundation under grant BSSGI0\_155816 `PlanetsInTime'.
G.-D.M. acknowledges the support of the DFG priority program SPP 1992 `Exploring the Diversity of Extrasolar Planets' (KU 2849/7-1).
Parts of this work have been carried out within the framework of the NCCR PlanetS supported by the Swiss National Science Foundation.
\end{acknowledgements}

\bibliographystyle{aa}
\bibliography{HIP65426.bib}

\end{document}